\documentclass{emulateapj}
\usepackage{apjfonts}
\usepackage{epsf}

\shortauthors{Bell et al.}
\shorttitle{Ten Gyr of quenching star formation in galaxies}

\slugcomment{{\sc The Astrophysical Journal}: in press }

\begin{document}



\title{What turns galaxies off?  The different morphologies of star-forming and quiescent galaxies since $z \sim 2$ from CANDELS  }


\author{Eric F.\ Bell$^1$, Arjen van der Wel$^2$, Casey Papovich$^3$, Dale Kocevski$^4$, Jennifer Lotz$^5$, Daniel H.\ McIntosh$^6$, Jeyhan Kartaltepe$^7$, S.\ M.\ Faber$^4$, Harry Ferguson$^5$, Anton Koekemoer$^5$, Norman Grogin$^5$, Stijn Wuyts$^8$, Edmond Cheung$^4$, Christopher J.\ Conselice$^9$, Avishai Dekel$^{10}$, James S.\ Dunlop$^{11}$, Mauro Giavalisco$^{12}$, Jessica Herrington$^1$, David C.\ Koo$^4$, Elizabeth J.\ McGrath$^4$, Duilia de Mello$^{13,14}$, Hans-Walter Rix$^2$, Aday R.\ Robaina$^{15}$, Christina C.\ Williams$^{12}$ }
\affil{$^1$ Department of Astronomy, University of Michigan, 500 Church St., Ann Arbor, MI 48109, USA; \texttt{ericbell@umich.edu} \\
$^2$ Max-Planck Institut f\"ur Astronomie, K\"onigstuhl 17, D-69117, Heidelberg, Germany \\
$^3$ George P.\ and Cynthia Woods Mitchell Institute for Fundamental Physics and Astronomy, and Department of Physics and Astronomy, Texas A\&M University, College Station, TX 77843-4242, USA \\
$^4$ UCO/Lick Observatory, Dept.\ of Astronomy and Astrophysics, University of California, Santa Cruz, CA 95064, USA \\
$^5$ Space Telescope Science Institute, 3700 San Martin Drive, Baltimore, MD 21218, USA \\
$^6$ Department of Physics, University of Missouri-Kansas City, Kansas City, MO 64110, USA \\
$^7$ NOAO-Tucson, 950 North Cherry Ave., Tucson, AZ 85719 \\
$^8$ Max-Planck-Institut f\"ur Extraterrestrische Physik, Giessenbachstrasse, D-85748 Garching, Germany \\
$^9$ University of Nottingham, School of Physics and Astronomy, Nottingham NG7 2RD, UK \\
$^{10}$ Racah Institute of Physics, The Hebrew University, Jerusalem 91904, Israel \\ 
$^{11}$ Institute for Astronomy, University of Edinburgh, Royal Observatory, Blackford Hill, Edinburgh EH9 3HJ, UK \\
$^{12}$ Department of Astronomy, University of Massachusetts, Amherst, MA 01003, USA \\
$^{13}$ Department of Physics, The Catholic University of America, Washington, D.C. 20064 \\
$^{14}$ Observational Cosmology Laboratory, Goddard Space Flight Center, Code 665, Greenbelt, MD 20771  \\
$^{15}$ Institut de Ciencies del Cosmos, ICC-UB, IEEC, Marti i Franques 1, 08028, Barcelona, Spain }

\begin{abstract}
We use {\it HST}/WFC3 imaging from the CANDELS Multicycle Treasury Survey, in conjunction with the Sloan Digital Sky Survey, to explore the evolution of galactic structure for galaxies with stellar masses $>3 \times 10^{10} M_{\sun}$ from $z=2.2$ to the present epoch, a time span of 10 Gyr.  We explore the relationship between rest-frame optical color, stellar mass, star formation activity and galaxy structure.  We confirm the dramatic increase from $z=2.2$ to the present day in the number density of non-star-forming galaxies above $3 \times 10^{10} M_{\sun}$ reported by others.  We further find that the vast majority of these quiescent systems have concentrated light profiles, as parameterized by the S\'ersic index, and the population of concentrated galaxies grows similarly rapidly.  We examine the joint distribution of star formation activity, S\'ersic index, stellar mass, inferred velocity dispersion, and stellar surface density.  Quiescence correlates poorly with stellar mass at all $z< 2.2$.  Quiescence correlates well with S\'ersic index at all redshifts.  Quiescence correlates well with `velocity dispersion' and stellar surface density at $z>1.3$, and somewhat less well at lower redshifts. Yet, there is significant scatter between quiescence and galaxy structure: while the vast majority of quiescent galaxies have prominent bulges, many of them have significant disks, and a number of bulge-dominated galaxies have significant star formation.  Noting the rarity of quiescent galaxies without prominent bulges, we argue that a prominent bulge (and perhaps, by association, a supermassive black hole) is an important condition for quenching star formation on galactic scales over the last 10\,Gyr, in qualitative agreement with the AGN feedback paradigm.  
\end{abstract}



\section{Introduction}
\label{sec:intro}

The last decade of study has brought into sharper focus the bimodality of the star formation histories of galaxies.  For star-forming galaxies alone, there is a relatively tight distribution of star formation rates (SFRs) at a given mass ($\sim 0.3$\,dex scatter, with a fraction of outliers to high SFR; e.g., \citealp{brinchmann04}, \citealp{salim07}), persisting out to $z>2$ \citep{noeske07_obs,wuyts11sfr}.  The red sequence has SFRs substantially below those expected for star forming galaxies (but often with some star formation; see, e.g., \citealp{yi05}).  We will call these `quiescent galaxies' in what follows.  The relative prominence of the two populations is a function of stellar mass, surface density, inferred velocity dispersion $M/R \propto \sigma^2$, and galaxy structure \citep[e.g.,][]{strateva01,kauf03_dens,blanton03prop,franx08,bell_disk08,vd11,wake_quench,cheung}.  This correlation between the structural properties of galaxies with their stellar populations is important: it signals that the processes that determine the structures of galaxies at least correlate, and perhaps are the same as, the processes that shape whether or not a galaxy has cold gas and star formation.  Furthermore, these two populations evolve in their relative importance: while the star-forming population has a stellar mass function that evolves slowly \citep{bundy05,borch06,bellcont,peng10,brammer11}, the quiescent galaxy stellar mass function evolves rapidly from $z \sim 2$ to the present day \citep[largely in normalization by factors of $\sim 10$, but with modest or no evolution in shape or `characteristic' mass $M^*$;][]{bell04c17,borch06,faber07,brown07,taylor09,peng10,dom11,brammer11}.  

A great deal of work, both theoretical and observational, has been carried out to try to better understand the drivers of the evolution of these two populations, particularly why some galaxies appear capable of shutting off their star formation while others are incapable of doing so.  In this study, we will focus on processes that can shut down star formation in galaxies which reside in the center of their own dark matter halo.  The clear effects of gas removal/starvation in dense environments (see, e.g., \citealp{vdw10_env}, \citealp{peng10}, \citealp{weinmann10}, \citealp{peng11}, \citealp{weinmann11} for recent discussions of this issue using survey datasets) are only a minor contributor to the evolution of `cosmic averaged' galaxy population, owing to the small number of galaxies inhabiting dense environments \citep{peng10,vdw10_env}.  Accordingly, we do not discuss the effect of environment in this paper (see, e.g., \citealp{peng10} for a careful discussion of the effects of environment as a function of cosmic time).  An important point is that models that include the growth of the dark matter framework, gas cooling, star formation and stellar feedback alone fail to predict a widespread population of non-star-forming galaxies \citep{benson03,cattaneo06,somerville08,dave11}; all galaxies are expected to accrete cold gas and form stars.  

A number of possible mechanisms have been proposed to keep a galaxy in the center of its own halo free of a significant cold gas content.   Noting the strong tendency of quiescent galaxies to have prominent (or dominant) bulge components, it has long been thought that merging plays an important role in determining their structure \citep[e.g.,][]{toomre,barnes92,hernquist93,naab06,hof10}.  There is a variety of evidence that is qualitatively (and in certain cases quantitatively) consistent with this picture: the approximate equality of the merger rate and the quiescent galaxy formation rate \citep[e.g.,][]{hopkins10,robaina10}; the detailed kinematic structure of early-type galaxies \citep{naab06,hof10}; and, the empirical association between relatively younger stellar populations and substructure (tidal tails, shells, asymmetries, etc.) in quiescent galaxies \citep{sch92,tal09,gyory10}. 

Largely in the merger context, the possibility that feedback from accretion onto a supermassive black hole may drive gas out of galaxies \citep{kauf00,springel05} or keep gas around galaxies from cooling \citep{croton06,somerville08} has been explored.  A wide array of observations are, at least at face value, qualitatively consistent with such a picture: e.g., galaxies with big bulges have big black holes \citep{mag98,gueltekin09}, low-redshift galaxies without prominent bulges cannot shut off their star formation on their own \citep{bell_disk08}, rapid large-scale winds are observed around post-starburst galaxies and quasars \citep{tremonti07,prochaska09}, and the energy measured in AGN-inflated outflow cavities in the hot gas atmosphere of groups and clusters of galaxies is approximately consistent with the energy required to offset cooling \citep{best06,fabian06}.  There are a number of other possible mechanisms for shutting off star formation, however, that are not related to feedback.  A few examples are: the heating of the gaseous halo through virialization of the gas content \citep{naab07,db07,khochfar07,johansson09grav}, changes in the mode of gas accretion onto galaxies as a function of dark matter halo mass \citep{keres05,dekel06,cattaneo06,birnboim07}, and the possibility of the growth of a stellar spheroid stabilizing a gas disk \citep{martig09}.

\subsection{The goal of this paper}

Given the range of possible mechanisms for shutting off star formation on galactic scales for galaxies in the center of their own halos (`centrals' hereafter), gathering empirical insight into the properties of non star-forming galaxies as a function of cosmic epoch can be helpful.  It has been argued that the key parameter that correlates with the paucity of star formation is stellar surface density \citep{kauf03_dens,franx08}, or possibly velocity dispersion (\citealp{wake_quench}; or roughly equivalently as $M/R$; \citealp{franx08}).  Yet, for a sample of low-redshift galaxies from the SDSS, \citet{bell_disk08} instead argues that S\'ersic index correlates much better with a lack of star formation for galaxies in the center of their own halo (that could not have been stripped of their gas by external influences), as non-star forming galaxies have uniformly high S\'ersic indices but a range of surface densities that overlap with star-forming galaxies (see also \citealp{cheung}).  Such an investigation of S\'ersic indices has not been carried out at $z\ga 1$ owing to a lack of large-scale near-IR {\it Hubble Space Telescope} ({\it HST}) imaging until recently (see, e.g., \citealp{kriek09}, \citealp{cameron10}, \citealp{szomoru11}, and \citealp{vd11} for early progress towards this goal at $z \sim 2$; \citealp{wuyts11} explores this in more depth).

A not inconsiderable challenge in achieving this goal is the definition of what constitutes star-forming or quiescent galaxies.  At a given stellar mass, the SFR of galaxies depends strongly on redshift, evolving by a factor of 5--10 by $z \sim 1$ and another factor of 4 or so out to $z \sim 2$ \citep{zheng07,noeske07_obs, dunne09,karim11}.  Noting that the 1$\sigma$ scatter in SFR at a given stellar mass for the vast majority of star-forming galaxies is $\sim 0.3$\,dex \citep{noeske07_obs}, one could choose to define a quiescent galaxy as one that has a SFR more than 1$\sigma$ below the star forming galaxy locus at the redshift of interest, and a star-forming galaxy as any galaxy forming stars at a higher rate.  An alternative approach is to separate galaxies by their optical--near-IR colors (e.g., $U-V$/$V-J$; \citealp{wuyts07}, \citealp{williams09}), where galaxies dominated by old stellar populations are distinguishable from star-forming galaxies with even substantial dust reddening (as used by e.g., \citealp{williams09} and \citealp{brammer11}; see also \citealp{patel12}).  In this paper, we adopt both techniques.  We note that a galaxy at $z \sim 2$ which is defined as quiescent according to these two criteria may have a SFR considerably in excess of almost all star-forming disk galaxies at the present day.  While this means that our sample of quiescent galaxies does not have identical properties across all redshifts, it does isolate a sample of galaxies with unusually low SFRs at that epoch given their stellar masses (confirmed by 24{\micron} stacking) --- one would like to understand why their SFRs are unusually low at that epoch.   

In this paper, we explore the evolution of the structures of galaxies as a function of redshift, and how they relate to the star formation activity in those galaxies.  We use new near-infrared imaging from the Wide Field Camera 3 (WFC3) on the {\it HST} taken as part of the Cosmic Assembly Near-IR Deep Extragalactic Legacy Survey (CANDELS) Multi-Cycle Treasury program \citep{grogin11,koe11}, focusing on the $0.6<z<2.2$ galaxy population in the UKIRT IR Deep Sky Survey (UKIDSS; \citealp{lawrence07}) Ultradeep Survey (UDS; \S \ref{sec:data}).  These data allow us to explore the structure of the $M_* > 3 \times 10^{10} M_{\Sun}$ galaxy population in the rest-frame optical to $z \la 2.2$.  We supplement this with data from the Sloan Digital Sky Survey Data Release 2 \citep{dr2} to connect with the properties of local galaxies.  We perform two basic analyses to explore the evolution of the galaxy population.  First, we explore the evolution of the galaxy population drawn from an `equivalent' constant comoving volume as a function of redshift to get a sense of how the star formation and structural properties of the galaxy population evolve with cosmic epoch (\S \ref{sec:evol}).  Second, we explore the relationship between the structural parameters of galaxies and their star formation activity using the full sample at each epoch (to maximize number statistics), in an attempt to understand which structural parameters best correlate with a lack of star formation activity (\S \ref{sec:param}). In what follows, we use Vega magnitudes for rest-frame colors, assume that every star ever formed does so according to a universally-applicable \citet{chabrier} stellar IMF, and assume H$_0 = 70\,{\rm km\,s^{-1}\,Mpc^{-1}}$, $\Omega_{m,0} = 0.3$ and
$\Omega_{\Lambda,0} = 0.7$. 

\section{Data}
\label{sec:data}

\subsection{UDS imaging data}
\label{sec:imaging}

Our sample definition and selection is based on the public $K$-band selected photometric and photometric redshift catalog produced by \citet{williams09}\footnote{We choose to use this public catalog instead of other proprietary catalogs to better facilitate comparison with previous works and to allow easier reproduction of the results.  We confirm that the results and conclusions do not significantly change if repeated with the redshifts, colors and stellar masses from the currently proprietary catalog of S.\ Wuyts et al.\ (in preparation).}.   We adopt these redshifts in this paper as the basis for conversion of apparent magnitudes into rest-frame colors, magnitudes and stellar masses, and the conversion of apparent to physical sizes.  The \citet{williams09} catalog uses $J$ and $K$-band data for the UKIDDS UDS Data Release 1 \citep{lawrence07,warren07} in conjunction with $B$, $R$, $i$ and $z$-band imaging from the SXDS \citep{furusawa08} and 3.6{\micron} and 4.5{\micron} IRAC imaging data taken as part of the SWIRE survey \citep{lonsdale03}.  Total fluxes were calculated using an elliptical \citet{kron} aperture, and observed-frame fluxes were calculated using a matched 1$\farcs$75 circular aperture on PSF-matched images (with the obvious exception of the poorer resolution IRAC imaging data, whose fluxes were measured in 3$\arcsec$ apertures and aperture corrected to the smaller aperture size).  Finally, \citet{williams09} used the EAZY photometric redshift code \citep{eazy} to estimate photometric redshift from the photometric catalogs.    

Spectroscopic redshifts for the UDS are relatively few in number and preferentially focus on brighter sources; for sources with $z_{\rm spec} \la 1.1$, the redshift normalized median absolute deviation (the median absolute deviation, renormalized to give the same value as the RMS of a Gaussian distribution) of $|z_{\rm phot}-z_{\rm spec}|/(1+z_{\rm spec})$ is $\sim 0.033$ with 8\% catastrophic outliers \citep{williams09}.  We have also compared the \citet{williams09} photometric redshifts against those of S.\ Wuyts et al.\ (in prep.), who used completely different (deeper) photometry and a similar photoz code to estimate photoz for galaxies in the CANDELS/UDS coverage, finding a $\Delta z/(1+z_{\rm Williams}) \sim 0.055$ and $\sim 10$\% catastrophic outliers (defined as having $|\Delta z|/(1+z_{\rm Williams}) > 0.2$.  We have confirmed that use of the photometric redshifts, $k$-corrections and stellar masses from Wuyts et al.\ instead of the public \citet{williams09} photometric redshifts plus the rest-frame colors and masses reported here yields no significant changes to our results.

To explore the structure and morphology of the $0.6<z<2.2$ galaxy population, we use near-infrared F160W imaging from {\it HST} using WFC3.  CANDELS is a {\it HST} Multi-Cycle Treasury program (PIs: S. Faber \& H. Ferguson, PID: GO-12060) to image five fields on the sky using the WFC3 and Advanced Camera for Surveys (ACS).  The CANDELS imaging of the UDS field includes 2/3 orbit in F125W and 4/3 orbits in F160W, split into two epochs (see \citealp{koe11} and \citealp{grogin11} for more details).  A total of 44 WFC3 tiles were imaged in the UDS; when these are cross-matched with the \citet{williams09} catalog, a total of $0.056$ square degrees is covered.  

We use also 24{\micron} flux as a diagnostic to aid in the separation between galaxies with active star formation and those with little or no star formation.  We use 24{\micron} public data from the SpUDS survey\footnote{http://irsa.ipac.caltech.edu/data/SPITZER/SpUDS } of the UDS.  PSF fitting photometry with a 13$\arcsec$ radius was performed, aperture corrected to total flux; the limiting flux of the catalog is $\sim 50 \mu$Jy at the $\sim 80\%$ completeness (or $4 \sigma$ level), and uncertainties in 24{\micron} flux are of order $\sim 20$\% (largely reflecting uncertainties in converting aperture to total flux, and source confusion).  

\subsection{Rest-frame quantities and stellar masses}
\label{sec:mass}

The rest-frame magnitudes and stellar masses used in this paper were calculated using a set of template spectral energy distributions from the P\'EGASE stellar population models (see \citealp{pegase} for a description of an earlier version of this stellar population model).  Such models give similar results to those of \citet{bc03} but stellar masses $\sim 0.15$\,dex higher than the models by \citet{maraston05}\footnote{Although note that \citet{kriek10} demonstrate that the \citet{bc03} models appear to fit the optical--near-IR SEDs of galaxies with large intermediate-age stellar populations substantially better than \citet{maraston05} models do. The overall mass scale offsets between \citet{bc03} and \citet{maraston05} models have been discussed by a number of papers, and recently by e.g., \citet{brammer11} and \citet{dom11}.}.  These stellar population templates have solar metallicity (as would be approximately expected for galaxies in the mass range $M_* > 3 \times 10^{10} M_{\sun}$; \citealp{gallazzi05}), and we fit for dust attenuation as a free parameter following \citet{calzetti01}, with values of gas $E(B-V)$ between $-0.05$ and 1.5 (to provide a little flexibility to fit negative attenuation to account for small photometry problems, etc.; in practice small negative attenuation values are rare in the fits presented here).  The templates include a broad range of exponentially-decreasing, constant or exponentially rising star formation histories, beginning at $z_f \sim 4$ (see \citealp{maraston10} for a discussion of the importance of exponentially rising star formation histories for fitting $z\ga 2$ SEDs).  

The templates treat the evolving galaxy population self-consistently, in the sense that all the galaxies that are in the $z=2$ template set also appear in, e.g., the $z=1$ or $z=0.6$ template sets further along their evolutionary path.  Thus, the galaxy population at lower redshifts is required to have substantial older stellar populations, driving up somewhat the typical age of the stars and the typical stellar mass-to-light ratio at a given color.  This is to be contrasted with other codes (e.g., \citealp{wuyts08} or \citealp{pannella09}) which have redshift-independent template sets, but exclude those templates that start star formation before the Big Bang --- the typical stellar populations of galaxies with this type of template fit end up being significantly younger than in the method discussed here.  It is unclear to us at this stage which method is more realistic; we will explore this issue with star formation histories drawn from a semi-analytic model of galaxy formation in a future paper (see, e.g., \citealp{lee10} for a version of this exercise at higher redshift).

These templates were compared with photometric data points of each galaxy, given the photometric redshift value from \citet{williams09}, and the template with the smallest $\chi^2$ was used to calculate rest-frame magnitudes and the stellar M/L ratio.  Rest-frame magnitudes were calculated by using the SED shape of the template to work out the predicted rest-frame magnitude of the object given the two nearest observed-frame bands, and then performing a weighted average of those two estimates of rest-frame magnitude.  Stellar masses were estimated using the stellar M/Ls of the best-fitting template, referenced to the three longest observed wavelengths for overall normalization.  We note that the inclusion of dust, while it improves the quality of SED fit, leads to only modest changes in the rest-frame colors and stellar masses of the sample explored in this paper, given the overall degeneracy between the effects of dust extinction and stellar population age in the optical--near--IR spectral region \citep{bdj}.  

\begin{figure}[t]
\begin{center}
\includegraphics[width=8.0cm]{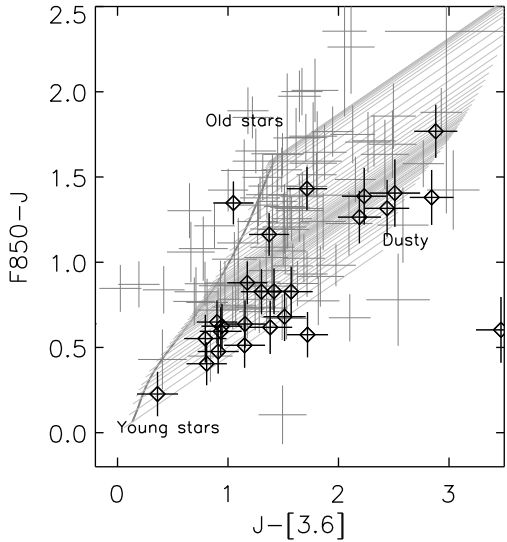}
\caption{Observed-frame color-color diagram for galaxies in the GOODS South field with spectroscopic (diamonds with black error bars) or photometric (thin grey error bars) redshifts in the range $1.4<z<1.6$.  Overplotted in gray are the colors of the stars-only template (thick gray line), dust extinction ($-0.05<E(B-V)<1.5$; included in the fitting, thin light gray lines).  The filters were chosen to have wavelengths similar to rest-frame $UVJ$ at the redshift of interest. 
\label{fig:colcol} 
}
\end{center}
\end{figure}

An illustration of this technique is given in Fig.\ \ref{fig:colcol}, where we show observed-frame colors for galaxies with inferred stellar masses in excess of $10^{10} M_{\sun}$ with spectroscopic (diamonds with black error bars) or photometric (thin grey error bars) redshifts in the GOODS South field (we use this field for this example, as it has many more spectroscopic redshifts in this range; the photometric redshift sample from UDS shows the same trends).  Overplotted with a thick gray line is the sequence of model composite stellar populations.  Thin light gray lines show dust attenuation vectors corresponding to $-0.05<E(B-V)<1.5$.  As described above, we fit these stellar population models with dust attenuation as a free parameter.  The vast majority of galaxies are well-explained by the template set; note that those few that are not covered completely by the template set are still assigned rest-frame magnitudes that reflect the observed magnitudes (and therefore also lie off the template rest-frame colors in a similar way), as the observed magnitudes are used to determine rest-frame magnitudes, in conjunction with a small template-dependent correction.  

This method, when applied to galaxies with independently-estimated stellar masses (from independent photometry of similar but not identical datasets) in this field (S.\ Wuyts et al., in prep.), the GOODS-S field \citep{wuyts08}, or the COSMOS field \citep{pannella09} yields similar masses for intensely star-forming galaxies and masses $\sim 0.2$ dex larger for more dusty star-forming galaxies or non-star-forming galaxies (as the templates used here are more dominated by older stellar populations than the those used by \citealp{wuyts08} or \citealp{pannella09}), with a scatter of 0.2 dex.  Rest-frame colors are reproduced to within 0.1 mag.  When this method is compared with the masses of \citet{bell03mf}, calculated on identical photometry and using a more restricted set of stellar population models without dust, there is no offset and 0.07 dex scatter in stellar masses, and a scatter of less than 0.05 mag in rest-frame $U-V$ colors.  

We adopt the stellar mass estimates described above for the purposes of this paper to ensure consistency of the stellar mass estimation method and stellar mass scale as a function of redshift (the same code was used to estimate stellar masses at all redshifts, and the choice of templates evolves consistently from redshift to redshift).  We have confirmed that the systematic discrepancies in stellar mass between the masses adopted here and those by, e.g., \citet{wuyts08} would operate to strengthen our conclusions (or in the case of the second part of the paper, leave them unaffected); their stellar masses for $z>0.6$ galaxies are systematically lower by up to a factor of 2, with a factor of 2 scatter, and the evolution of the population would appear more rapid than it appears in this paper.  

In what follows we adopt a mass limit of $3 \times 10^{10} M_{\sun}$; the sample is `complete' above this limit.  Completeness, often meant to signify the limit above which no galaxies are missing as a result of selection, is not straightforward to calculate for multi-band photometric redshift surveys, as the magnitude limits in the different optical/near--IR bands limit the recovery of photometric redshifts and stellar masses in ways that are spectral type and redshift-dependent.  \citet{williams09,williams10} argue the 5$\sigma$ $K$-band survey limit corresponds to a stellar mass limit of $\log(M_{5\sigma} / M_{\sun}) \sim 10.2$; they analyze their sample to a limit of  $\log(M_{\rm Williams} / M_{\sun}) \sim 10.6$ to ensure accurate UDS-derived galaxy size estimates (for which higher $S/N$ is required).  We choose to analyze the sample to $\log(M_{lim} / M_{\sun}) = 10.5$.  Given that completeness is such a challenge to calulate, we have tested this limit empirically by repeating the analyses in this paper on an independent, currently proprietary set of magnitudes, photometric redshifts, rest-frame magnitudes and stellar masses drawn from deeper imaging data (S.\ Wuyts et al., in prep.), finding that our results and conclusions do not significantly change.  The masses and rest-frame magnitudes of the sample presented in this paper are available for download at: http://www.astro.lsa.umich.edu/$\sim$ericbell/data.php 

\subsection{S\'ersic profile fits of $0.6<z<2.2$ galaxies}

To describe the structure of the galaxy population at $0.6<z<2.2$, we use parametric \citet{sersic} fits to the galaxy images (A.\ van der Wel et al., in preparation).  A surface brightness profile of the form $\Sigma(r) = \Sigma_e exp[-\kappa(\frac{r}{r_e}^{1/n}-1)]$ is fit using the GALFIT package \citep{galfit}, and the GALAPAGOS wrapper \citep{galapagos}, allowing the magnitude, axis ratio $b/a$, position angle, half-light radius $r_e$, S\'ersic index $n$, and central position to be free parameters.  GALAPAGOS estimates the sky value explicitly on larger scales, leading to more robust fits given typical image cutout sizes (see \citealp{boris07} for a detailed discussion of testing of our method using both simulated and deeper data; A.\ van der Wel in prep.\ demonstrate that the uncertainty in fit parameters caused by sky estimation errors in this particular dataset are substantially smaller than the uncertainties that we adopt below from comparison of independent S\'ersic fits).  The S\'ersic index $n$ describes the shape of the light
profile, where $n=1$ corresponds to an exponential light profile
and $n=4$ corresponds to a $r^{1/4}$ law profile characteristic
of massive, spheroid-dominated early-type galaxies.  The S\'ersic parameter $n$ is a reasonably good proxy for the ratio of bulge luminosity to total luminosity, as illustrated in Fig.\ 14 of \citet{simard11} --- systems with high $n$ invariably host a prominent bulge, whereas systems with low $n$ host a weak or no bulge component. 
At the depths typical 
of this imaging, uncertainties in the fit parameters are $\delta \log_{10} n \sim 0.15$\,dex, $\delta r_e \sim 18$\%, and $\delta b/a \sim 0.07$, as constrained from both fits of simulated galaxies and independent GALFITs to F125W imaging of a subsample of $z \sim 1.6$ galaxies in the UDS \citep{papovich11}. 

The S\'ersic fits adopted in this paper are carried out on the F160W imaging data of CANDELS.  This corresponds to rest-frame wavelength ranges of $\lambda_{\rm rest} \sim 0.55/0.65/0.9${\micron} for $z \sim 2/1.4/0.8$ systems.  A possible concern is that this change in rest-frame wavelength may affect the demographics of the population.  While the full dataset at shorter wavelengths in the UDS field has not been analyzed with GALFIT, it is possible to test if this may be an issue using GALFIT on a smaller set of CANDELS F160W data in the GOODS South area, in comparison with published GALFITs on the F850LP ACS GEMS data for the extended Chandra Deep Field South \citep{boris07}.  We choose galaxies with $0.4<z<0.8$ for this test, where the F850LP data span the same range in rest-frame wavelength as the F160W data for $1.3 < z < 2.2$.  For systems with low $n \la 1$, we find a slight tendency for the F160W S\'ersic index to exceed the F850LP data (by $\la 0.1$ dex), and for $n \ga 2$ there is no systematic difference between the two sets of S\'ersic fits.  The scatter around these modest offsets is $\sim$0.2\,dex, equivalent to the combined uncertainties of the fits.  The fraction of systems with $n > 2.5$ in F850LP (restframe $\sim 0.6${\micron}) is in fact 20\% {\it larger} than the fraction derived using the F160W imaging (restframe $\sim 1${\micron}).  This indicates that the fraction of $n>2.5$ galaxies presented here at $0.6 \la z \la 1.3$ is likely to be close to, or perhaps up to 20\% lower than the evolution of the $n>2.5$ fraction if it were measured in the rest frame: this operates to make the evolution of the population demographics {\it more rapid still} than we measure in a fixed (red) observed band. We conclude that our use of F160W data alone across the $0.6<z<2.2$ redshift range is not an important source of systematic error in this analysis.

\subsection{SDSS parameters for the low-redshift comparison sample}

In order to connect with the present-day galaxy population, we use a sample
of low-redshift galaxies explored in \citet{bell_disk08} from the 
SDSS Data Release Two \citep{dr2} and presented in the NYU Value-Added Galaxy Catalog \citep{vagc}.  
We use foreground extinction-corrected \citep{sfb98}, 
$k$-corrected \citep{bell03mf}
$r$-band absolute Petrosian magnitude for the
galaxy absolute magnitude (random and systematic 
uncertainties $\la 0.15$\,mag), and model
colors for higher S/N estimates of galaxy color 
(uncertainties $\la 0.05$\,mag).
Following \citet{bell03mf}, we have merged this catalog with the Two Micron All-Sky Survey (2MASS; \citealp{2mass}) to facilitate SED fitting and to allow splitting of galaxies by $U-V$/$V-J$ into quiescent and star-forming populations.   Stellar masses and rest-frame colors were estimated from $ugrizJK$ photometry using the above stellar population model templates.  

Star formation and AGN 
classifications and estimates of total star formation rate
were taken from \citet{brinchmann04} using emission 
line measurements described in \citet{tremonti04}.  Galaxies are classified
as star forming, AGN, composites, or are left unclassified (typically
because the galaxies lack line emission in their SDSS spectra).  

The only source of S\'ersic fits for all galaxies in our sample is \citet{blanton03prop}\footnote{\citet{simard11} also fit all galaxies in the SDSS with single S\'ersic profile fits.  Owing to the low redshift of our SDSS sample, about 1/8 of the galaxies in this particular sample lack fits in \citet{simard11}, as they are above the bright limit adopted for the construction of those catalogs. Accordingly, we do not adopt the estimates of \citet{simard11} for this work. The vast majority of galaxies that do have fits agree well with the trends reported below between NYU and S\'ersic fits performed directly to the imaging data (put differently, corrected NYU fits agree with \citealp{simard11}).  Results would be unchanged if we adopted those estimates instead for the fraction of galaxies with fits in \citet{simard11}.   }. 
\citet{blanton03prop}
fit the light profile of galaxies in the SDSS, measured in circular apertures, with a seeing-convolved
\citet{sersic} profile for all of the galaxies in the VAGC.  The S\'ersic fits by \citet{blanton03prop} give values for the S\'ersic index, in particular, that are offset from $n$ values determined using 2D galaxy image fits.  We have compared the S\'ersic indices (and other fit parameters) from \citet{blanton03prop} to fits carried out by \citet{vdw08} on a small subset of galaxies in the NYU VAGC.  We find that the S\'ersic index estimates are related: $\log_{10} n_{\rm 2D} \sim -0.39 + 1.75 \log_{10} n_{\rm NYU\,VAGC}$, with 0.2 dex scatter.  Half-light radii show the following correlation: $\log_{10} r_{\rm 2D} \sim \log_{10} r_{\rm NYU\,VAGC} - 0.05 + 0.025 n_{\rm 2D,new}$, with 0.1 dex scatter, where $n_{\rm 2D,new}$ is the estimate of equivalent 2D S\'ersic index derived from the NYU VAGC S\'ersic index using the above relation.  A similar analysis was carried out with completely independent GALFITs by \citet{guo09}, and importantly the above trends are identical in the case of $n$, and $\la 0.05$\,dex different in the case of $r$, to the median offsets as a function of $n$ in their Fig.\ A1.  Recall that the 0.2\,dex scatter between the `rescaled' NYU $n$ values and those of \citet{vdw08} or \citet{guo09} (or the 0.1\,dex scatter in radii) is comparable to the typical joint uncertainties in any comparison of even unbiased values of $n$ (or $r$).  We conclude that these rescaled NYU $n$ and $r$ values are unbiased, have uncertainties comparable to those determined directly from 2D fits, and are appropriate for connecting the evolution of $n$ and $r$ with the results of 2D fitting for the $z>0.6$ galaxies.

\section{The evolving relationships between stellar mass, color and morphology}
\label{sec:evol}

\subsection{Separating galaxies into quiescent and star forming populations}

\label{sec:sep}

\begin{figure*}[t]
\begin{center}
\includegraphics[width=16.0cm]{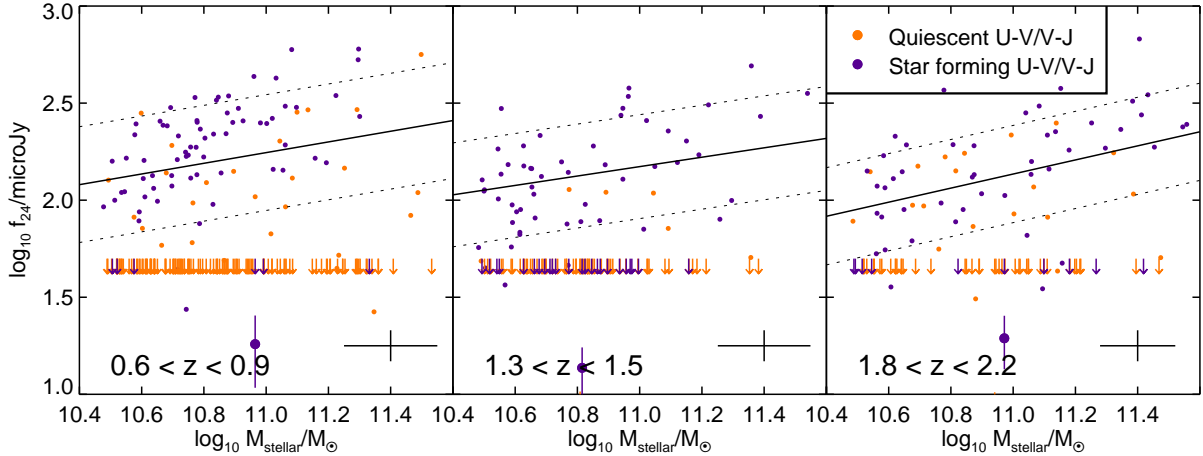}
\caption{Flux at 24{\micron} as a function of stellar mass in three different redshift bins; detections are shown as filled symbols and $4 \sigma$ upper limits shown by arrows.  The thick solid line and dotted lines are the robust linear fits to the data and the $\pm 1 \sigma$ range.  Colors denote classification according to $U-V$/$V-J$ (Fig.\ \protect\ref{fig:uvvj}): purple symbols denote galaxies that lie in the star forming part of the color--color diagram, and orange galaxies lie in the quiescent region of the plot.  The large solid purple points with error bars show the stacked 24{\micron} flux and uncertainty for galaxies individually undetected at 24{\micron} but in the star-forming part of the $U-V$/$V-J$ diagram.  
\label{fig:24um} 
}
\end{center}
\end{figure*}

\begin{figure}[t]
\begin{center}
\includegraphics[width=8.0cm]{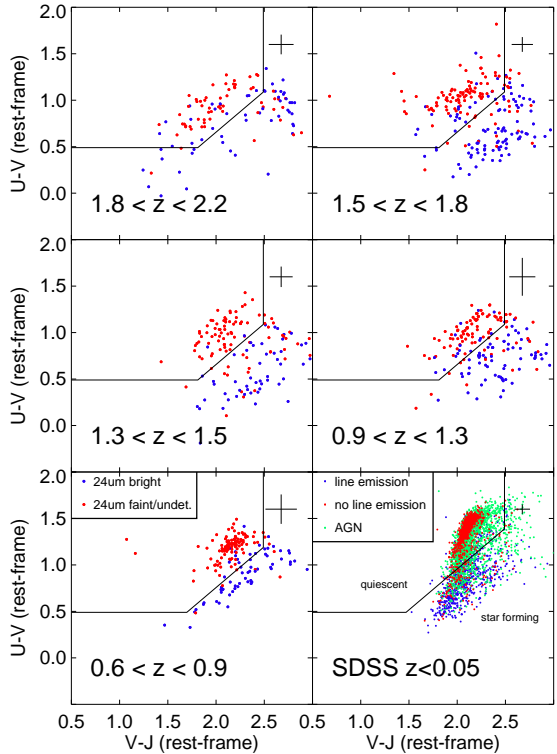}
\caption{Rest-frame $U-V$ as a function of $V-J$ color for the six redshift bins used in this paper.  Galaxies are color-coded according to their 24{\micron} properties ($z>0.6$) or emission line properties ($z<0.05$); (quiescent) galaxies not detected or fainter than $-1\sigma$ from the star forming galaxies locus are color-coded red, and the rest of the galaxies (all star forming) are color coded blue.  Green galaxies at $z<0.05$ are galaxies in the SDSS with AGN-like emission lines.  Superimposed are the rest-frame color cuts used in this paper, following \citet{williams09}. All galaxies have stellar masses in excess of $3 \times 10^{10} M_{\sun}$.  
\label{fig:uvvj} 
}
\end{center}
\end{figure}

\begin{figure*}[t]
\begin{center}
\includegraphics[width=13.0cm]{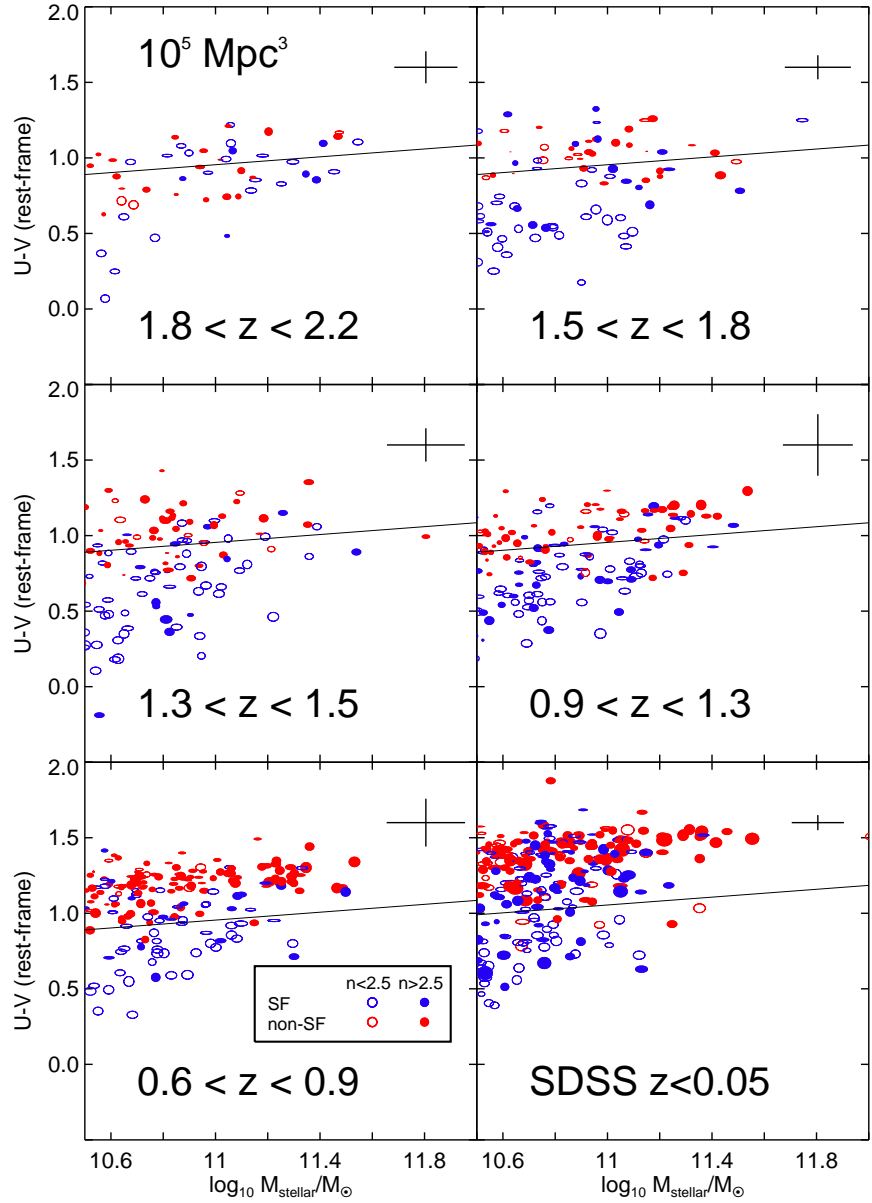}
\caption{{\small The evolution of $U-V$ rest-frame color (in Vega magnitudes) as a function of stellar mass, in six different redshift bins.  Galaxies that appear in this figure are subsampled to a fixed comoving volume of $10^5 {\rm Mpc}^3$ at all redshifts to show the evolution of the massive galaxy content of a `representative' volume with cosmic time.  Open symbols show galaxies with $n < 2.5$ and filled symbols show galaxies with $n>2.5$.  For all symbols the axis ratio of the galaxy is reflected by the axis ratio of the symbol, and the size of the symbol scales with $(1+\log_{10} r_e/{\rm kpc})$.  The black line is shown in all panels at the approximate locus of $z \sim 2$ red galaxies.  Galaxies are color-coded by star formation activity --- galaxies classified as quiescent by both 24{\micron}/SFR and $U-V$/$V-J$ diagnostics are color-coded red, and blue symbols show all other galaxies.  AGN with $z<0.05$ are classified using only $U-V$/$V-J$.  }
\label{fig:colmass} 
}
\end{center}
\end{figure*}

\begin{figure*}[t]
\begin{center}
\includegraphics[width=13.0cm]{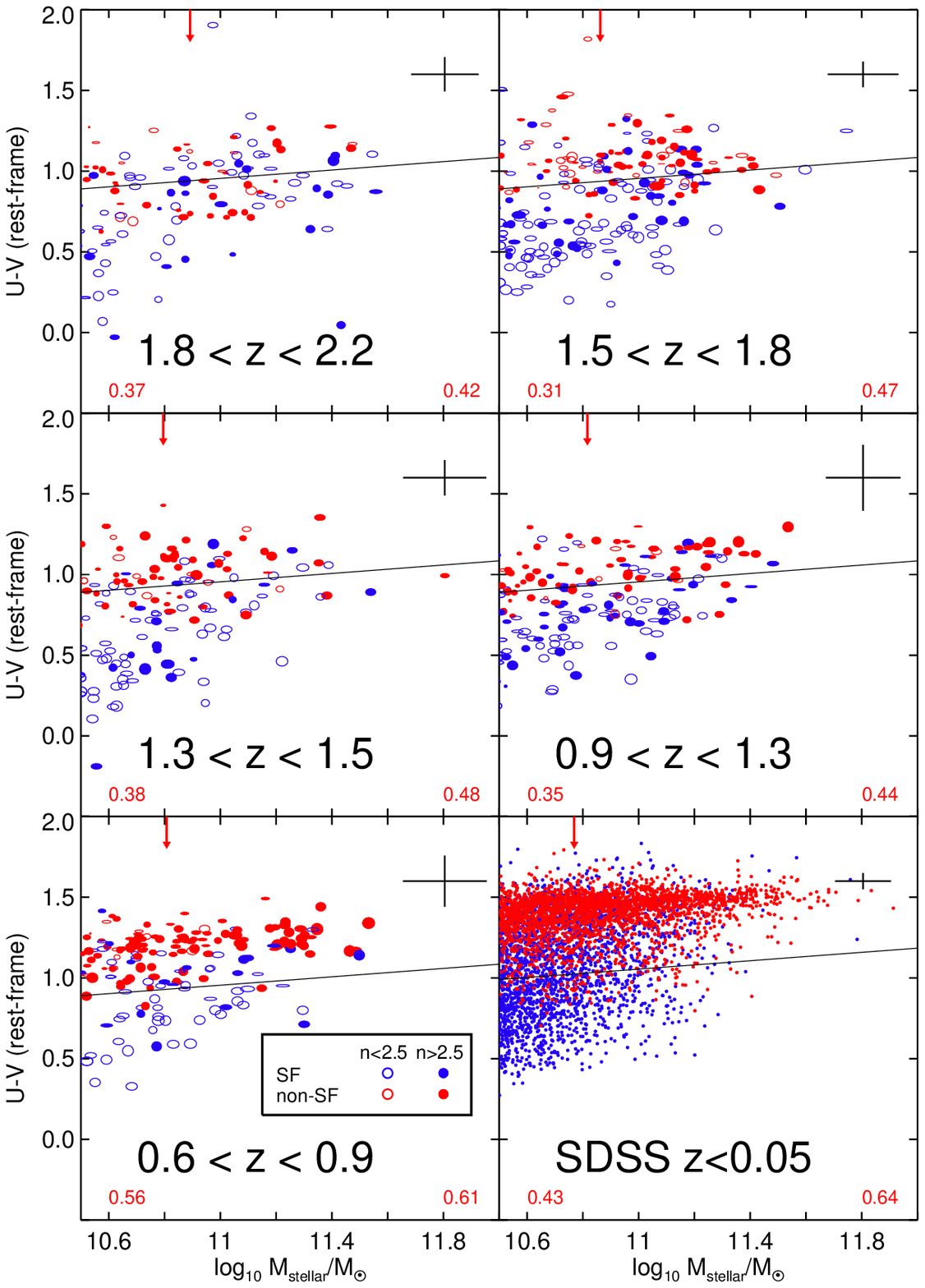}
\caption{{\small The evolution of $U-V$ rest-frame color (in Vega magnitudes) as a function of stellar mass, in six different redshift bins.  Symbols are as in Fig.\ \protect\ref{fig:colmass} for $z>0.6$.  For the SDSS, we code galaxies only by their star formation activity for clarity.  The first five panels $(z>0.6)$ are for the full UDS survey and the SDSS panel uses the whole SDSS DR2 subsample used in \protect\citet{bell_disk08}.  In all panels, the small red numbers on the left/right show the fractions of quiescent galaxies in the halves of the sample below and above the median mass (red arrow), respectively.}
\label{fig:colmass_all} 
}
\end{center}
\end{figure*}

We separate galaxies by their star formation activity using two sets of independent diagnostics: mid infrared (or, for the SDSS, emission line) information, and position on optical--near-infrared color-color diagrams.  We use both cuts in this paper; we describe the 24{\micron}-derived cuts first (showing how they relate to rest-frame color cuts) and then show the rest-frame color cuts (showing how they relate to 24{\micron} cuts).   

At $z>0.6$, galaxies are classed in large part according to their 24{\micron} emission properties as an admittedly imperfect proxy for obscured SFR (we use emission-line diagnostics and SFR estimates from \citealp{brinchmann04} for galaxies from the SDSS).  Substantial 24{\micron} flux can also result from AGN activity. We do not attempt to discriminate between AGN activity and SF activity for our purposes here, simply noting that systems with 24{\micron} flux dominated by AGN at this 24{\micron} luminosity and redshift range are not the dominant population \citep[e.g.,][]{donley08,kartaltepe10} and that we are primarily attempting to weed out galaxies whose rest-frame optical colors are a poor reflection of the stellar populations in that galaxy, a goal for which our simple approach is sufficient.  

In this spirit, we wish to avoid an explicit, and uncertain, conversion of 24{\micron} flux into SFR \citep[e.g.,][]{papovich07,elbaz11,wuyts11sfr}.  At each redshift of interest, we fit the relationship between 24{\micron} flux and stellar mass for galaxies detected at 24{\micron} (shown as filled symbols), as shown in Fig.\ \ref{fig:24um}.  The approximate trend at all redshifts is $\log_{10} f_{24}/\mu{\rm Jy} \sim 2 + 0.5 \log_{10} (M_*/3\times 10^{10} M_{\sun})$, with a scatter of less than 0.3\,dex (we use the actual fits and scatters, which vary slightly with redshift, to perform the split into star forming and quiescent).  The slope and scatter of this relationship is well-documented and studied \citep[e.g.,][]{salim05,zheng07,noeske07_obs,karim11}; it is a remarkable coincidence that the zero point in terms of 24{\micron} flux varies so little with redshift, owing to the interplay between the dramatic reduction of SFR at a given stellar mass with decreasing redshift, the luminosity distance and the redshift-dependent 24{\micron} $k$-correction.  Quiescent galaxies then must have a 24{\micron} flux (UDS) or SFR (SDSS) lower than $-1 \sigma$ from the star forming galaxy locus.

In Fig.\ \ref{fig:24um}, we have color-coded symbols by their position on the rest-frame $U-V$/$V-J$ diagram (Fig.\ \ref{fig:uvvj}), using the slightly redshift-dependent cuts described in \citet{williams09}.  Orange symbols show galaxies with rest-frame optical--near-IR colors characteristic of quiescent galaxies, and purple symbols show galaxies with colors characteristic of star-forming galaxies with a range of reddening values.  In Fig.\ \ref{fig:uvvj}, we show the optical--near-IR colors of galaxies in the six redshift intervals of interest.  In contrast to Fig.\ \ref{fig:24um}, we have color-coded the symbols in Fig.\ \ref{fig:uvvj} by 24{\micron} flux (UDS) or SFR (SDSS).  Galaxies with 24{\micron} fluxes/SFRs lower than $-1 \sigma$ from the star forming galaxy locus have been color-coded red, and galaxies with fluxes/SFRs higher than $-1 \sigma$ from the SF galaxy locus have been color-coded blue.  In the $z<0.05$ slice, emission-line diagnostics are available, and any object with AGN-like lines or composite SF/AGN lines \citep{brinchmann04} has been color-coded green.

Inspection of Figs.\ \ref{fig:24um} and \ref{fig:uvvj} shows the large degree of overlap, and the complementarity of having both explicit 24{\micron}/SFR information and $U-V$/$V-J$ colors (see also \citealp{williams09}, \citealp{wuyts09}, \citealp{brammer11}, and a morphological investigation by \citealp{patel12}).  Galaxies with quiescent $U-V$/$V-J$ tend, for the most part, to be undetected at 24{\micron} (Fig.\ \ref{fig:24um}).  There are exceptions to this: a few galaxies with quiescent $U-V$/$V-J$ at $z \la 1.5$ are detected at 24{\micron}, and by $z \sim 2$ it is clear that for our particular dataset the contamination of the quiescent region of $U-V$/$V-J$ color space by 24{\micron}-detected objects is significant.  Stacking at 24{\micron} of the remaining individually-undetected galaxies with quiescent $U-V$/$V-J$ yields marginal (2-3$\sigma$ significance) detections at the $5-10 \mu$Jy level at all redshifts (indicating SF/AGN activity a factor of $>10$ lower than typical star-forming galaxies at that redshift; see also \citealp{papovich06} for an early discussion of star formation in red-selected galaxies; these measurements are not shown as they fall off of the range of data values plotted).

Conversely, galaxies detected clearly at 24{\micron} are almost always in the star forming region of $U-V$/$V-J$ (Fig.\ \ref{fig:uvvj}), but again with some exceptions (e.g., at $z \sim 1.4$ there is a clear group of galaxies with star forming colors that are individually undetected at 24{\micron}).  Stacks of those few individually 24{\micron} undetected galaxies with star-forming colors yields significant detections at the 15-25$\mu$Jy level (blue filled points with error bars on Fig.\ \ref{fig:24um}), a factor of a few below the SF galaxies locus, largely consistent with an interpretation of these systems as the low SFR tail of the SF galaxy population.  This high degree of correspondence between the two methods has been shown before by e.g., \citet{williams09} and \citet{wuyts09}.  Fig.\ \ref{fig:uvvj} also shows that the use of rest-frame color information for $z<0.05$ is particularly valuable; galaxies classified as AGN can have either quiescent or star forming colors.

\begin{figure*}[t]
\begin{center}
\includegraphics[width=13.0cm]{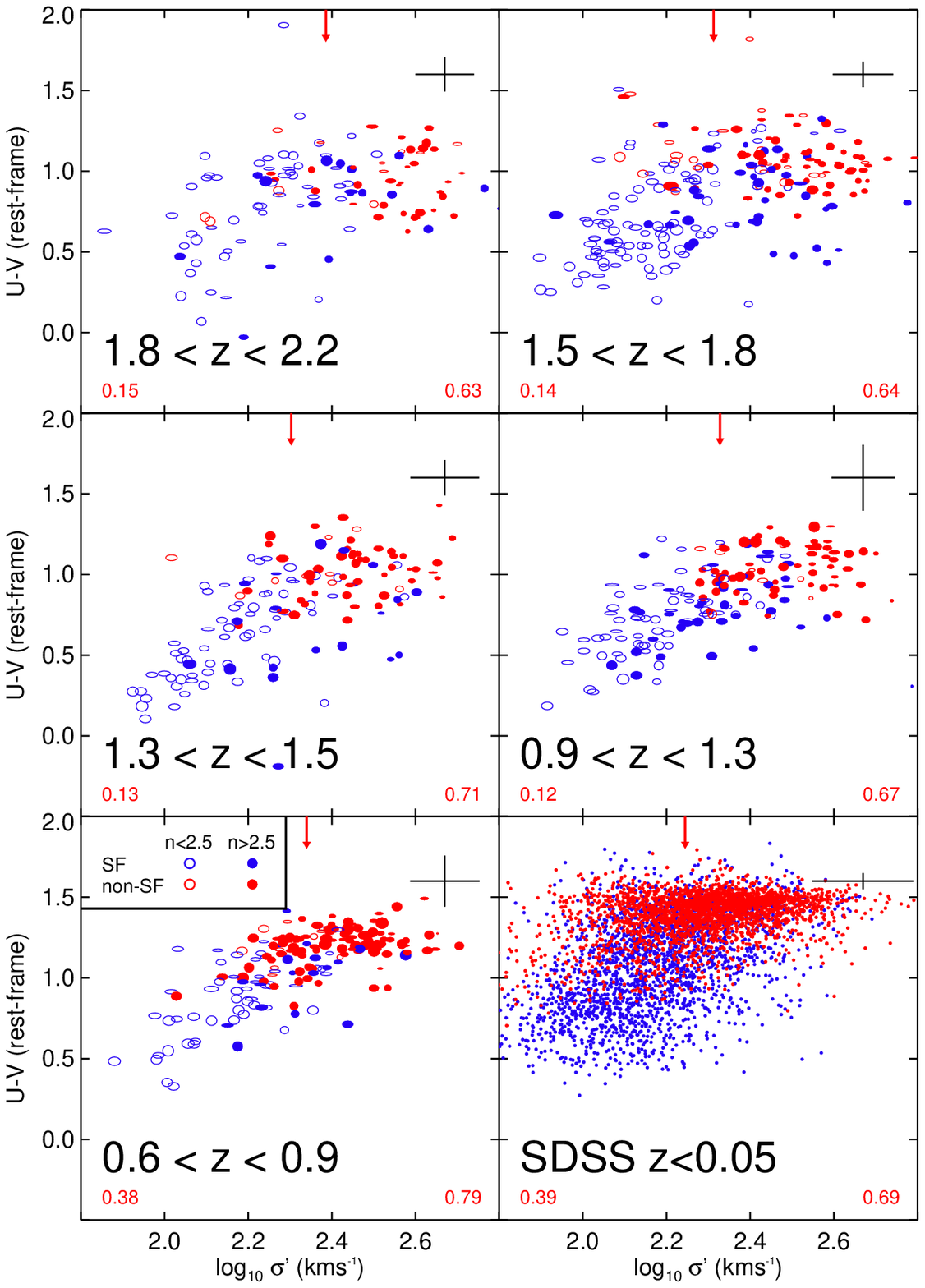}
\caption{The evolution of $U-V$ rest-frame color (in Vega magnitudes) as a function of estimated velocity dispersion $\sigma^{\prime}$ (scaling as $(M_*/r_e)^{1/2}$ with a S\'ersic index-dependent proportionality constant), in six different redshift bins.  Symbols are as in Fig.\ \protect\ref{fig:colmass} for $z>0.6$.  For the SDSS, we code galaxies only by their star formation activity for clarity.  The first five panels $(z>0.6)$ are for the full UDS survey and the SDSS panels use the whole SDSS DR2 subsample used in \protect\citet{bell_disk08}.  In all panels, the small red numbers on the left/right show the fractions of quiescent galaxies in the halves of the sample below and above the median $\sigma^{\prime}$ (red arrow), respectively.
\label{fig:colsig} 
}
\end{center}
\end{figure*}

\begin{figure*}[t]
\begin{center}
\includegraphics[width=13.0cm]{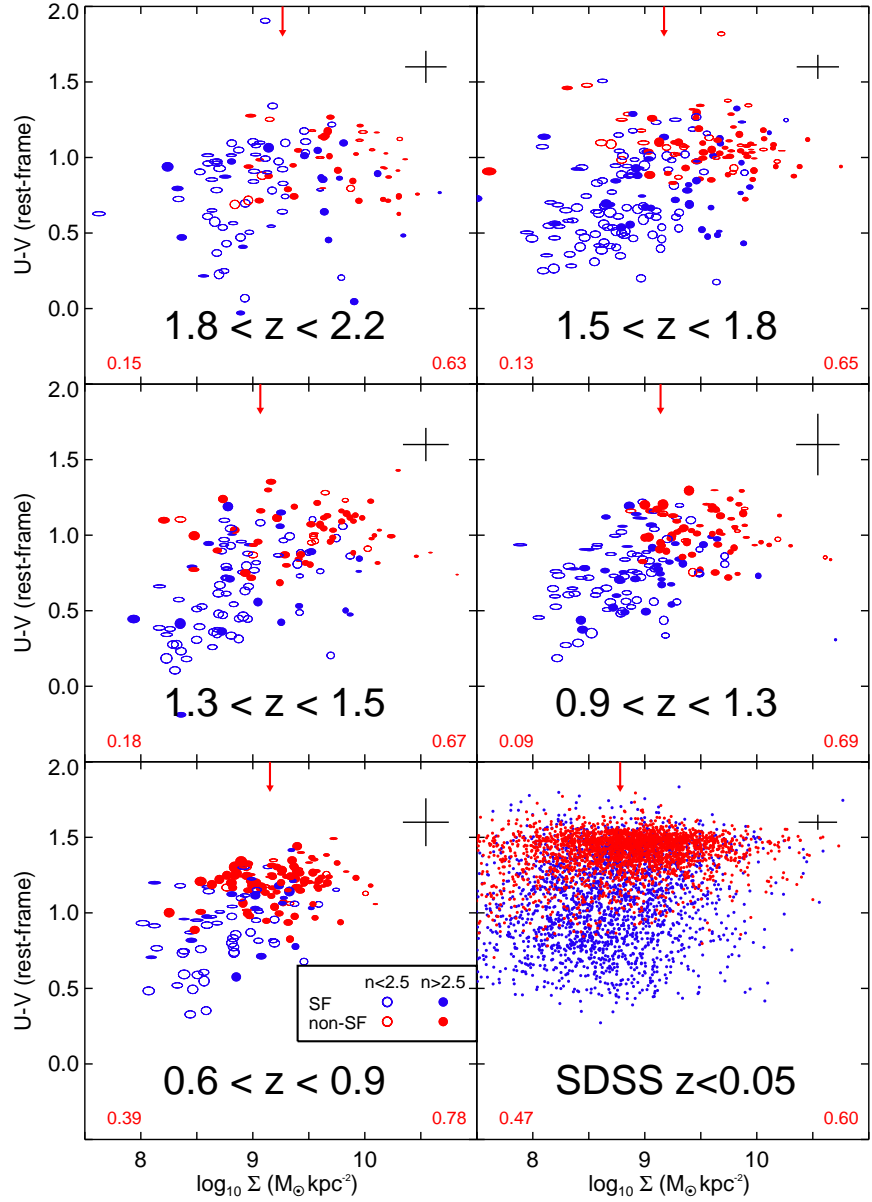}
\caption{The evolution of $U-V$ rest-frame color (in Vega magnitudes) as a function of stellar surface density $0.5M/ \pi r_e^2$, in six different redshift bins. The samples and symbols are as in Fig.\ \protect\ref{fig:colsig}.   In all panels, the small red numbers on the left/right show the fractions of quiescent galaxies in the halves of the sample below and above the median stellar surface density (red arrow), respectively. 
\label{fig:coldens} 
}
\end{center}
\end{figure*}

\begin{figure*}[t]
\begin{center}
\epsscale{0.7}
\includegraphics[width=13.0cm]{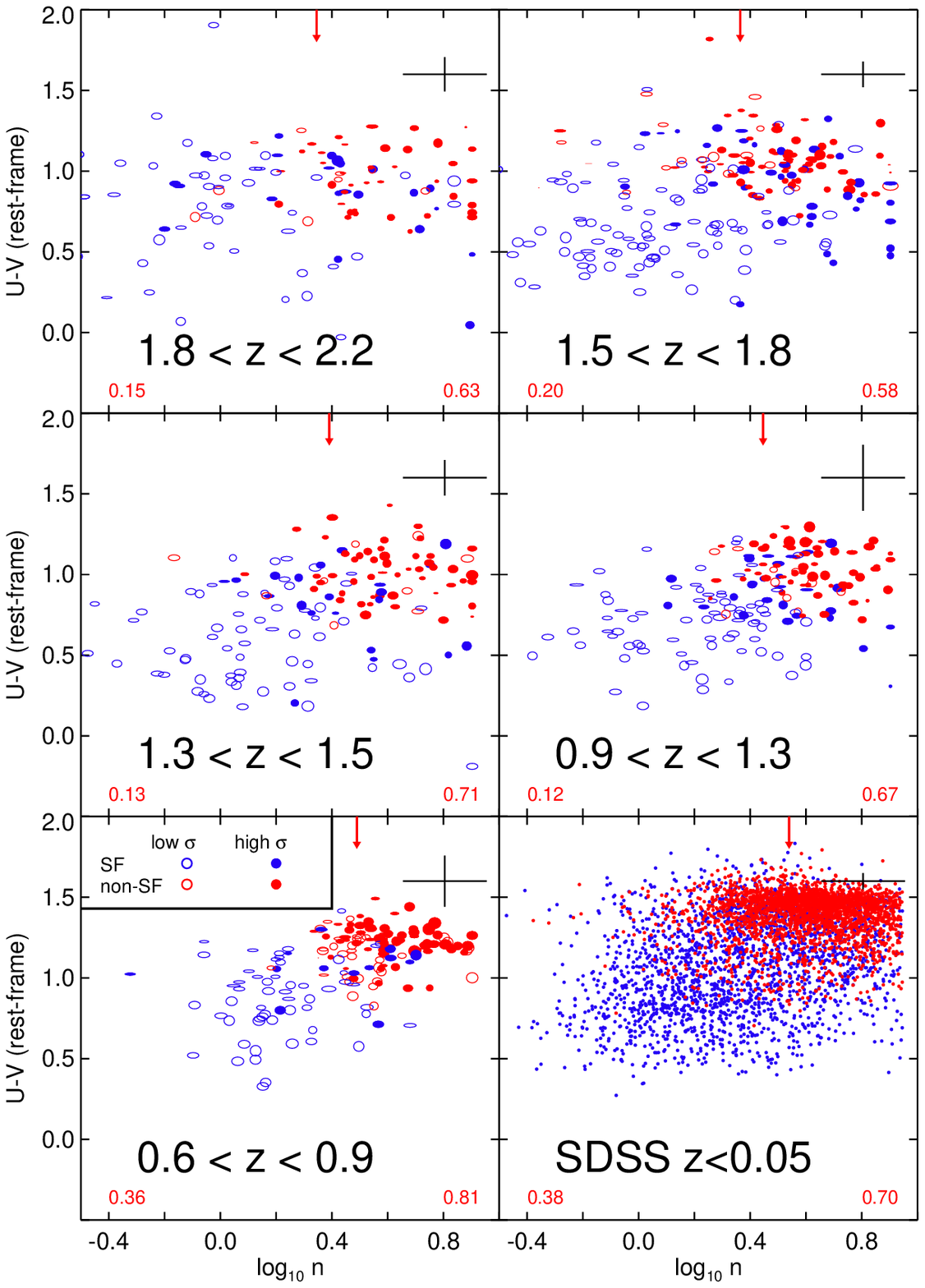}
\caption{The evolution of $U-V$ rest-frame color (in Vega magnitudes) as a function of S\'ersic index, in six different redshift bins.  The samples and symbols are as in Fig.\ \protect\ref{fig:colsig}, except that now filled/open symbols denote those galaxies with above/below median $\sigma^{\prime}$ at that redshift (in the stellar mass range considered in this paper).   In all panels, the small red numbers on the left/right show the fractions of quiescent galaxies in the halves of the sample below and above the median S\'ersic index (red arrow), respectively.
\label{fig:colser} 
}
\end{center}
\end{figure*}

We separate galaxies into quiescent and star forming using both criteria to capitalize on their different strengths and shortcomings.  Discrimination by $U-V$/$V-J$ is sensitive to lower amounts of star formation than 24{\micron} separation, especially at $z \sim 1.4$ and $z \sim 2$.  On the other hand, separation by 24{\micron} is considerably less sensitive to photoz error than $U-V$/$V-J$, as one simply needs to know which redshift bin the galaxy is in, and even some cross-talk between bins can be tolerated.  We define quiescent galaxies as having {\it both} `quiescent' colors in $U-V$/$V-J$ and 24{\micron} fluxes/SFRs lower than $-1 \sigma$ from the star forming galaxy locus at the redshift of interest.  For SDSS galaxies with AGN-like emission lines, we split only on the basis of $U-V$/$V-J$.  Star forming galaxies are defined as those which satisfy either (or both) of the $U-V$/$V-J$ star forming galaxy color cuts or having 24{\micron} fluxes/SFRs brighter than $-1 \sigma$ from the star forming galaxy locus.   
One can see that incorporating 24{\micron} data and insight from $U-V$/$V-J$ into this analysis is crucial.  As can be seen directly in Figs.\ \ref{fig:24um} and \ref{fig:uvvj}, and further appreciated by the intermixing of blue and red symbols at red rest-frame $U-V$ in Figs.\ \ref{fig:colmass}--\ref{fig:colser}, failure to flag galaxies by multiwavelength-derived SF activity leads to considerable confusion between star forming galaxies with substantial dust columns and non star-forming galaxies, especially at higher redshifts \citep{taylor09,brammer11}.  

\subsection{Evolution of the galaxy population in a given comoving volume}

Fig.\ \ref{fig:colmass} shows the rest-frame $U-V$ colors of galaxies, as a function of their stellar mass, in six different redshift bins.  We choose to show the properties of the galaxy population as a function of $U-V$ rest-frame color to connect with other studies \citep[e.g.,][]{bell04c17,borch06,ruhland09,whitaker10,brammer11} and as a joint (rather sensitive) constraint on SFH and dust content.  Galaxies are color-coded by star formation activity (\S \ref{sec:sep}): red symbols show galaxies classified as quiescent using {\it both} 24{\micron} information and $U-V$/$V-J$ colors, and blue symbols show the remaining population.  In all panels of this figure, the galaxy population has been Monte Carlo subsampled down to an equivalent comoving volume of $10^5 {\rm Mpc}^3$ by adjusting the number of galaxies to track the number density of galaxies with $M_* > 3 \times 10^{10} M_{\sun}$ determined from larger surveys (the line in Fig.\ \ref{fig:ngal}).  Put differently, variations in the number of galaxies from panel to panel illustrate true evolution in the galaxy population (as the volume is fixed; see Appendix \ref{ap:norm} for further discussion).  Filled symbols show galaxies with $n > 2.5$.  Open symbols show galaxies with $n<2.5$.  In all panels, the linear size of the symbol scales with $(1 + \log_{10} r_e/{\rm kpc})$, where $r_e$ is the half-light semi-major axis, and the axis ratio of the symbol is the same as that of the galaxy of interest. The black line is shown in all panels for reference at the approximate position of $z \sim 2$ non star-forming galaxies.  

The evolution of the galaxy population in the epoch $z \sim 2$ to the present day is obvious.  As has been argued by a number of other authors \citep[e.g.,][]{arnouts07,fontana09,taylor09,ilbert10,cassata11,dom11,brammer11}, there is dramatic evolution in the overall number of galaxies with $M_* > 3 \times 10^{10} M_{\sun}$ (as quantified in Fig.\ \ref{fig:ngal}).  Furthermore, Fig.\ \ref{fig:colmass} shows that the evolution of the number density of quiescent galaxies is particularly striking (again, as has been argued by the above cited works).  Fig. \ref{fig:colmass} makes it clear however that the evolution of the star formation activity of the intermediate-mass and massive galaxy population is accompanied by a large-scale change in the structure of galaxies (see also \citealp{wuyts11}, who see similar behavior) from a $z \ga 1.5$ population dominated by low $n$ (little or no bulge), mostly star-forming systems to the present population, dominated by galaxies with high $n$ (with a prominent bulge), many of them quiescent (but not all of them).  Quantitatively, there are a factor of $\sim $2.5 (3.5) more quiescent ($n>2.5$) galaxies with $M_* > 3 \times 10^{10} M_{\sun}$ today than there were {\it galaxies} with those masses at $z \sim 2$, respectively.  Thinking about it differently, the current population of quiescent ($n>2.5$) galaxies with $M_* > 3 \times 10^{10} M_{\sun}$ is approximately as numerous as the entire $M_* > 3 \times 10^{10} M_{\sun}$ population at $z \sim 1.1$ (0.7) respectively.  This change in global demographics from $z \sim 2$ to the present day makes it clear that, in addition to processes that shut off star formation on galactic scales, there must also be (the same or different) processes that lead to an associated change in the surface brightness profiles of galaxies over the same time period (and given the correspondence between a lack of star formation and structure, the time scales of such processes must be comparable).  

Fig.\ \ref{fig:colmass} shows also that the scatter in the quiescent galaxy color--magnitude relation (CMR) decreases towards lower redshift.  The evolution of CMR scatter from $z \sim 2$ to the present day is well-documented in the literature \citep{ruhland09,whitaker10}.  The scatter in the U-V colors of non star-forming galaxies with $U-V > 0.6$ at $z \ge 1.3$ is 0.17 mag (our measurement), very consistent with the carefully-measured results of \citet{whitaker10}, who find a scatter of 0.13-0.2 mag for $1.3\le z < 2$.  \citet{ruhland09} find that the scatter in $U-V$ color at $z \la 1$ is $\sim 0.1$ mag (measured much more carefully than the CMRs presented in Fig.\ \ref{fig:colmass}; our measurements also give a scatter of $\sim 0.1$ mag), essentially independent of redshift.  Modeling presented in both \citet[for $z<1$]{ruhland09} and \citet[at $1<z<2$]{whitaker10} shows that the evolution of CMR scatter is naturally interpreted as being caused by a constant inflow of new galaxies onto the red sequence at the observed number density growth rate.

\section{Empirical correlations between a lack of star formation and galaxy structure}
\label{sec:param}

\begin{figure*}[t]
\begin{center}
\includegraphics[width=18.0cm]{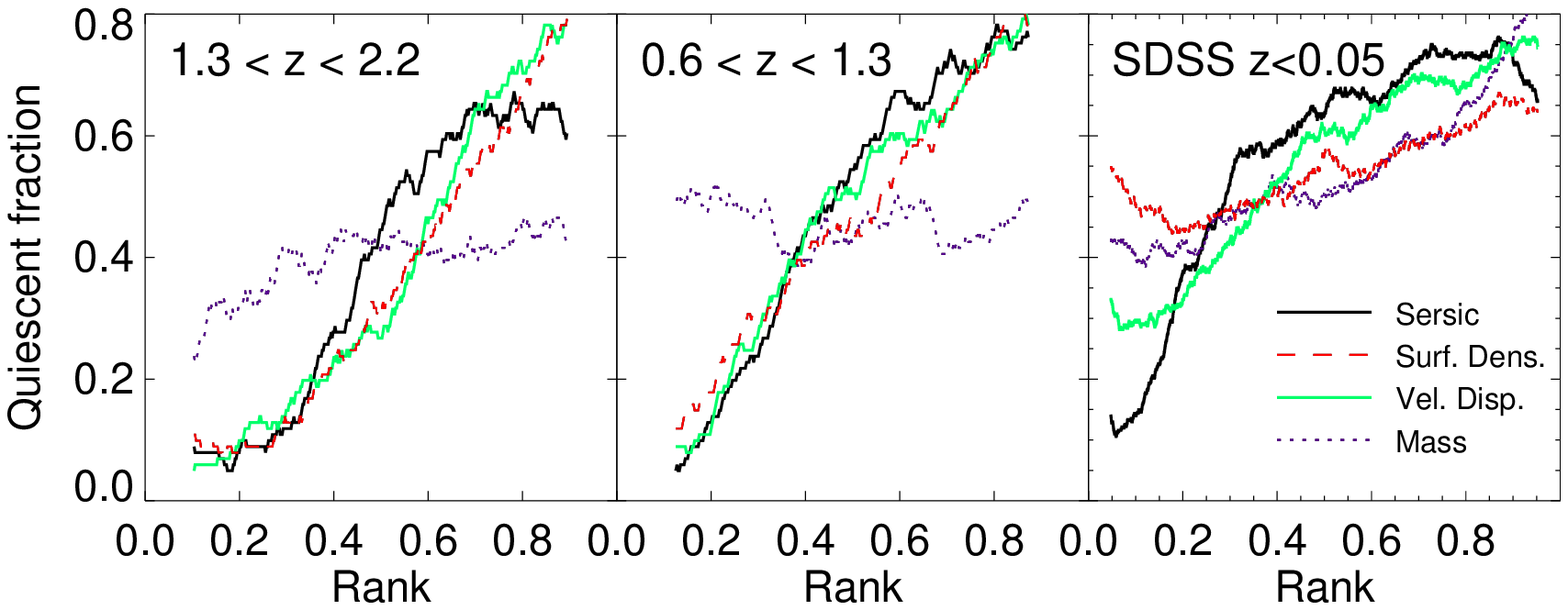}
\caption{The quiescent fraction in three broad redshift bins as a function of the rank of $n$, $\Sigma$, $\sigma^{\prime}$, and $M_*$ (rank denoting where a galaxy is in the sorted list of the $n$, $\sigma^{\prime}$ or $M_*$ in the sample at that redshift of interest).  These are running average fractions determined for the $\pm$50 (250 for the SDSS) nearest neighbors in rank of the quantity in question.  Mass is always a poor predictor of quiescence; S\'ersic index is clearly superior to $\Sigma$ or $\sigma^{\prime}$ at $z<0.05$, performs as a slightly better predictor of quiescence at $0.6<z<1.3$, and performs as well as $\sigma^{\prime}$ or $\Sigma$ at $1.3<z<2.2$.  
\label{fig:hist} 
}
\end{center}
\end{figure*}

\begin{figure*}[t]
\begin{center}
\includegraphics[width=17.0cm]{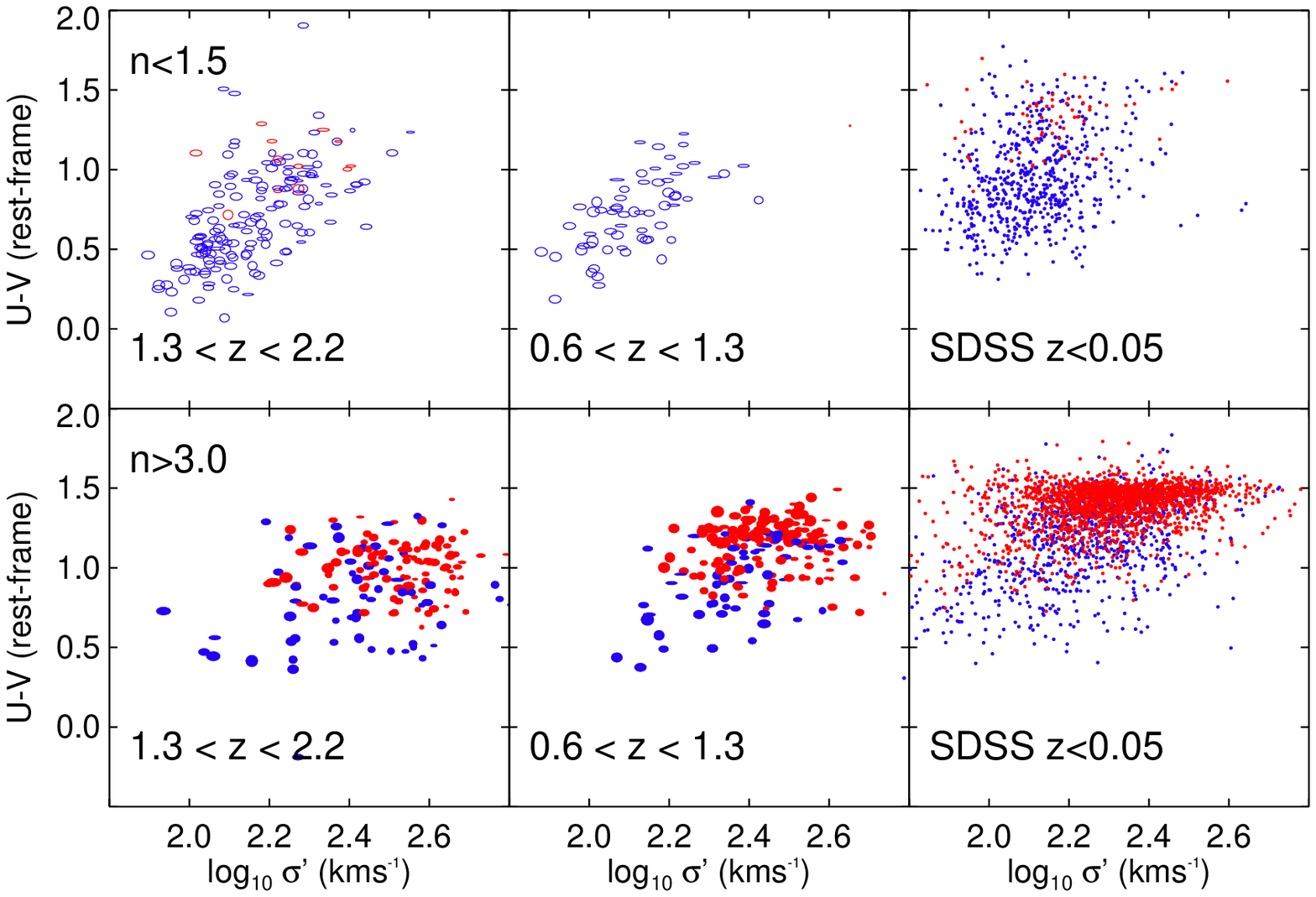}
\caption{The evolution of $U-V$ rest-frame color (in Vega magnitudes) as a function of $\sigma^{\prime}$ in three broad redshift bins.  The top panels show galaxies with $n<1.5$ (galaxies with little or no bulge), the bottom panels show only galaxies with $n>3$ (galaxies with a prominent bulge component).  Symbols are as in Fig.\ \ref{fig:colsig}.   
\label{fig:colsep} 
}
\end{center}
\end{figure*}

\subsection{Broad trends}

One of the most notable trends seen by \citet{arnouts07}, \citet{taylor09}, \citet{ilbert10}, \citet{dom11}, \citet{brammer11} and elaborated upon in Fig.\ \ref{fig:colmass} is the dramatic growth of the quiescent galaxy population from $z \sim 2$ to the present day.  Fig.\ \ref{fig:colmass} demonstrates also that with the growth of the quiescent galaxy population comes a concurrent growth of the population of concentrated $n>2.5$ galaxies (see also \citealp{wuyts11}).   \citet{franx08} argues that these galaxies also have high surface density and $M/r_e$, which should scale with velocity dispersion, and that $M/r_e$ is the parameter that best correlates with a lack of SF activity at $z \la 2$ \citep[see also][]{wake_quench,cheung}.  In this section, we explore how the different structural parameters correlate with star formation activity in an attempt to gain possible insight into the processes that drive galaxies into quiescence.  

In Fig.\ \ref{fig:colmass_all} shows the color--mass trends using the full UDS and SDSS DR2 datasets, as opposed to sub-sampling down to an equivalent volume of $10^5 {\rm Mpc^3}$, in order to delineate the trends with better fidelity than the subsample shown in Fig.\ \ref{fig:colmass}.  

In Fig.\ \ref{fig:colsig}, we show rest-frame color varies with velocity dispersion estimated from stellar mass, half-light radius and S\'ersic index, where: $\sigma^{\prime 2} = \frac{GM_*}{K_{\nu}(n) r_e \sqrt(b/a)}$, and $K_{\nu}(n) = 0.954+\frac{73.32}{10.465+(n-0.94)^2}$ scales $M/r_e$ in a physically-motivated way to account for the structure of a galaxy via the S\'ersic index \citep[amounting to velocity dispersions at a fixed $M/r_e$ that are $\sim 0.1$ dex lower for $n \la 2$ systems compared to those with $n \sim 4$]{bertin02}.  Such a scaling permits recovery of observed velocity dispersions of galaxies in the SDSS as a function of photometric parameters to an accuracy of $\sim 0.12$\,dex (see, e.g., \citealp{taylor10}, \citealp{bezanson11}).  This velocity dispersion estimate scales also with rotation velocity for rotationally-dominated systems, albeit with a different proportionality constant.   

In Fig.\ \ref{fig:coldens}, we show the variation in rest-frame color with surface density within the half-light radius $\Sigma = 0.5M/\pi r_e^2$ (following \citealp{kauf03_dens} and \citealp{franx08}, and as explored for low redshift by \citealp{bell_disk08}).  In all figures, symbols are coded by S\'ersic index, axis ratio, size and star formation activity as in Fig.\ \ref{fig:colmass}, with the exception of the SDSS sample where the number of galaxies allows color-coding by star formation activity alone.  

In Figs.\ \ref{fig:colmass}---\ref{fig:coldens}, one can see trends previously reported by \citet{kauf03_dens}, \citet{franx08}, \citet{vd11} or many subsequent studies: galaxies with high stellar mass, high velocity dispersion, or high surface density, tend not to form stars (where the latter study is particularly relevant owing to its use of {\it HST}-derived structural parameters and the equivalent width in H$\alpha$ as a star formation indicator).  Yet, one can see also evidence that $M_*$, $\sigma^{\prime}$ or $\Sigma$ fail to give a complete picture of which galaxies are quiescent (see also \citealp{bell_disk08}, \citealp{cheung}, \citealp{wake_quench}).  A significant fraction of low stellar mass galaxies are quiescent (therefore stellar mass is a relatively poor predictor of quenching), and a small fraction of galaxies with $z \la 1.5$ and intermediate or low values of $\sigma^{\prime}$ and $\Sigma$ are quiescent\footnote{The actual fraction of galaxies forming stars at low $\Sigma$ or $\sigma^{\prime}$ may be rather higher, as the sample is limited by stellar mass.  Fig.\ \ref{fig:colmass} shows that quiescent galaxies are smaller at a given stellar mass than star-forming galaxies, therefore it is possible that if Figs.\ \ref{fig:colsig} and \ref{fig:coldens} were $\sigma^{\prime}$ or $\Sigma$ limited samples they would show a more prominent population of (lower stellar mass) quiescent galaxies with relatively low $\sigma^{\prime}$ or $\Sigma$.}.  

Fig.\ \ref{fig:colser} shows the trend in rest-frame color with S\'ersic index (recall that S\'ersic index correlates with the relative prominence of a bulge component; \citealp{simard11}).  At $z>1$, this trend was not explored by \citet{franx08}, as they did not analyze large-scale {\it HST} imaging, and therefore lacked reliable measurements of surface brightness profile shape; this can, however, be compared with (and is consistent with) Fig.\ 1 of \citet{wuyts11}.  Symbols are similar to previous figures except that filled/open symbols now denote galaxies with above/below the median $\sigma^{\prime}$ at that redshift for the galaxies with $M_* > 10^{10.5} M_{\sun}$ in our sample (recall that the previous filled/open distinction was by S\'ersic index, which would be redundant).  Focusing on the $z>0.6$ points, one can see that galaxies with high S\'ersic index are much more likely to be non star-forming than their low S\'ersic index counterparts.  Furthermore, one can see that there is a range of velocity dispersions at a given S\'ersic index.  A similar qualitative behavior is seen for the SDSS galaxies, notwithstanding quantitative differences in the definition of S\'ersic index for the NYU VAGC catalog.   

\subsection{Which parameter correlates best with a lack of star formation?}

Figs.\ \ref{fig:colmass}--\ref{fig:colser} demonstrate that `typical' quiescent galaxies have higher mass, `velocity dispersion', surface density and S\'ersic index than `typical' star-forming galaxies.  In this section, we explore further which parameter correlates the best with a lack of star formation activity.

Fig.\ \ref{fig:hist} shows the quiescent fraction (evaluated in running bins of 101 galaxies at $z>0.6$, or 501 galaxies at $z<0.05$) as a function of the rank of a galaxy in stellar mass (dotted line), `velocity dispersion' (grey solid line), surface density (dashed line) and S\'ersic index (black solid line) in three broad redshift bins.  At all redshifts, stellar mass is a poor predictor of quiescence.  At $z<0.05$, S\'ersic index is clearly a better predictor of quiescence than any other parameter; in particular, galaxies with low $n$ overwhelmingly host detectable star formation.  

We quantify this by introducing the quantity $\Delta_{0.2-0.6}$, which quantifies the fractional difference in rank between the galaxy population being 20\% quiescent and 60\% quiescent (i.e., the rank difference corresponding to when the lines cross quiescent fractions of 0.2 and 0.6).  At $z<0.05$, $\Delta_{0.2-0.6}$ is undefined for stellar mass, surface density and velocity dispersion (as the quiescent fractions never go below 0.2), and is 0.31$\pm$0.03 for galaxies when ordered by S\'ersic index.  Put differently, when one orders the galaxies by S\'ersic index from low to high, from the point when the quiescent fraction is 0.2, one needs to go through 31\% of the galaxies to reach the point where the quiescent fraction reaches 0.6, and that bootstrapping of the galaxies being used to calculate this quantity leads to a $\pm$3\% variation of $\Delta_{0.2-0.6}$.  

At $0.6<z<2.2$, one can see a rather different situation: our estimate of velocity dispersion, surface density, and S\'ersic index all correlate comparably well with SF activity (the first and last trends agree with Fig.\ 4 of \citealp{vd11}).  For $0.6<z<1.3$, $\Delta_{0.2-0.6}$ is 0.31$\pm0.06$, 0.40$\pm0.09$ and 0.47$\pm 0.07$ for $n$, $\sigma^{\prime}$ and $\Sigma$ respectively.  For $1.3<z<2.2$, $\Delta_{0.2-0.6}$ is 0.30$\pm0.07$, 0.30$\pm0.06$ and 0.37$\pm 0.06$ for $n$, $\sigma^{\prime}$ and $\Sigma$ respectively.  Echoing in a muted way the behavior of the SDSS sample, one can see that the S\'ersic index still correlates with quiescence well (and as well as $z<0.05$; see also \citealp{vd11}). The correlation of star formation activity with $\sigma^{\prime}$ is as strong as that of S\'ersic index at $1.3<z<2.2$, and appears to weaken with decreasing redshift. The correlation of star formation activity with $\Sigma$ is marginally poorer than with S\'ersic index at $z>0.6$.   

The small numbers on the lower left and right hand corners of each panel of Figs.\ \ref{fig:colmass_all}--\ref{fig:colser} also help to illustrate this point.  These numbers show the fraction of quiescent galaxies for two different subsamples: the half with lowest stellar $M_*$, $\sigma^{\prime}$, $\Sigma$, or $n$, and the half with highest $M_*$, $\sigma^{\prime}$, $\Sigma$, or $n$.  We note that the quiescent fractions split by mass are quantitatively similar to those in Fig.\ 4 of \citet{brammer11}.  The S\'ersic index is the metric that tends maximizes the contrast between the two halves of the sample (except at $0.9<z<1.3$ and $1.5<z<1.8$, where S\'ersic index still correlates very well with quiescence).  Repeating these analysis with alternate photometric redshifts, stellar masses and rest-frame properties (S.\ Wuyts et al., in prep.) produces very minor changes, with a slightly stronger preference for $n$ as the parameter that best correlates with quiescence.

Given that $n$ and $\sigma^{\prime}$ correlate well with quiescence, it is interesting to explore star formation activity as a function of both parameters.  In the same three broad redshift bins, we show the $U-V$ color of galaxies as a function of $\sigma^{\prime}$.  Symbols are coded as they are in Figs.\ \ref{fig:colmass}--\ref{fig:coldens}.  The top panels show galaxies with $n < 1.5$ (galaxies with little or no bulge) and the bottom panels show galaxies with $n>3$ (galaxies with a prominent bulge).  The population with $n<1.5$ is overwhelmingly star forming.  In strong contrast, the $n>3$ population has a large quiescent fraction.  There is a strong tendency for high $n$ galaxies to have high $\sigma^{\prime}$ (although there is considerable scatter in velocity dispersion at a given S\'ersic index), and not all high velocity dispersion or S\'ersic galaxies lack star formation.  Neither parameter perfectly predicts quiescence, although it is clear that both perform very well at $z \ga 1$, and S\'ersic predicts quiescence better for local samples. 

There are two emergent themes that we wish to draw the reader's attention to.
First, Figs.\ \ref{fig:colser} and \ref{fig:colsep} show that, with very few exceptions, galaxies with low S\'ersic index all appear to form stars at all $z \la 2.2$.  The threshold appears to be somewhere around $n \sim 1.5-2$: at $n \la 2$, the fraction of quiescent galaxies is $\la 10$\% (and in many redshift bins it is less than a few percent).  When investigated in more detail using the same SDSS sample \citep{bell_disk08}, it was found that i) real low $n$ quiescent galaxies are all satellite galaxies in galaxy clusters, i.e., they are stripped disk galaxies, and ii) the few quiescent `low $n$' systems in the centers of their own halos that remained were in fact the result of measurement error in $n$, as visual inspection showed a distinct bulge component.  We show examples of some $n<2$ quiescent galaxies at $z>1.5$ in Fig.\ \ref{fig:nsfstmp} (at $0.9<z<1.5$ there are only three $n<2$ quiescent systems, and those look similar to the $z>1.5$ examples; the $0.6<z<0.9$ points are bad S\'ersic fits).  While some systems are relatively extended and have low $n$, and the one inclined galaxy is clearly reminiscent of a disk, most appear spheroidal and compact.  Given Figs.\ \ref{fig:colser} and \ref{fig:nsfstmp} in concert, it is clear the vast majority of quiescent galaxies have a prominent spheroid.  This extends the results of \citet{bell_disk08} determined for nearby galaxies and \citet{cheung} for $z\sim 0.65$ to $z \sim 2.2$, when the Universe was $\sim 1/4$ of its present age: galaxies lacking a prominent bulge appear to have great difficulty shutting off their own star formation on galactic scales.  

Second, Figs. \ref{fig:colser} and \ref{fig:colsep} make it clear that having a high S\'ersic index alone (or indeed, having high $n$, $\sigma^{\prime}$, $\Sigma$ and $M_*$) is not enough to ensure a lack of star formation.  At all redshifts, a small minority of high $n$ sources form stars at an appreciable rate.  This illustrates a key point of this paper; it appears that for all $z \la 2.2$ a large bulge is {\it necessary} to stop star formation, but is not {\it sufficient} to stop star formation.  This extends the conclusion of \citet{bell_disk08} determined for local galaxies out to $z \sim 2.2$, when the non star-forming galaxy population was considerably less prominent. 

\section{Discussion}
\label{sec:disc}

There are two main observational results in this paper: the rapid growth of the 
quiescent galaxy population between $z=2$ and the present day, and the recognition that this growth appears to be intimately linked to the growth of galaxies with prominent bulges (as quantified by high S\'ersic index, inferred velocity dispersion and surface density).   

\subsection{Musings on the mechanisms that prevent significant cold gas in galaxies}

These results have some bearing on understanding which mechanisms lead to quenching of star formation in galaxies.  Recall that the role of environmental quenching is relatively minor in our `cosmic averaged' population evolution, and that we are focusing on which types of physical process lead to quenching of star formation in galaxies in the centers of their halos (the mass quenching of \citealp{peng10}).  To facilitate this, we will set up two straw person hypotheses: suppression of star formation by feedback (either star formation or AGN feedback; `feedback quenching', e.g., \citealp{kauf00}, \citealp{croton06}, \citealp{somerville08}), or suppression of star formation because the halo reaches a certain critical mass (`halo quenching'; e.g., \citealp{dekel06}, \citealp{cattaneo06}).  

Fig.\ \ref{fig:hist} demonstrates that internal properties (S\'ersic index, and at $z \ga 1$ also `velocity dispersion' and surface density) correlate strongly with quiescence, whereas galaxy mass only weakly correlates with star formation activity.  Galaxy mass is expected to correlate well with halo mass.  \citet{more09} measure the scatter in luminosity at a given halo mass to be 0.16\,dex, and \citet{yang09} measure the scatter in stellar mass at a given halo mass to be 0.17\,dex, both substantially less than the $\sim 1$dex dynamic range probed in this work.  The observed weakness of correlation between quiescent fraction with stellar mass, coupled with the expected modest scatter between stellar mass (via its proxy, luminosity) and halo mass, implies a weak correlation between halo mass and quiescence at all $z<2.2$.   This is consistent with the claim by \citet{more11} that at fixed stellar mass there is {\it no difference} between the average halo masses of quiescent and star-forming central galaxies.  On this basis, it would be natural to conclude that quiescence is not determined by halo mass {\it alone}.

We caution that such a conclusion may be premature.  \citet{wake_halo} find that at high velocity dispersion/stellar masses the clustering of galaxies (a reflection of characteristic halo mass) is better described as a function of velocity dispersion than stellar mass. On this basis, our finding that quiescent fraction is quite strongly correlated with velocity dispersion may suggest that quenching is a strong function of halo mass.  We note, however, that the sample studied by \citet{wake_halo} is complete only at high mass, $\ga 10^{11} M_{\sun}$, where \citet{more09} infer increasing scatter in halo mass as luminosity increases.  Taking \citet{more09}, \citet{yang09} and \citet{wake_halo} together, it appears possible that luminosity correlates well with halo mass at lower luminosity/stellar mass and that velocity dispersion correlates better with halo mass at higher luminosity/stellar mass.  This issue cannot be resolved here, and the issue of how the weakness of the quiescence--stellar mass correlation should be interpreted by necessity remains open.

The results here demonstrate that systems with high S\'ersic index, `velocity dispersion' and high surface density are much more likely to be quiescent (see also \citealp{franx08}, \citealp{vd11}, \citealp{wuyts11}, \citealp{wake_quench}).  All of these metrics correlate strongly with the relative prominence of a bulge component.  Given correlations between black hole mass, bulge mass, velocity dispersion and S\'ersic index \citep{haering04,peng06,graham07,gueltekin09}, our results tentatively support the AGN feedback paradigm, at least at the qualitative level.  In particular, the seeming inability of galaxies with low S\'ersic indices/inferred velocity dispersion/surface density to shut off their star formation is very naturally interpreted in this framework \citep{bell_disk08} --- no supermassive black hole, no shut-off of star formation.  We note that feedback from star formation-driven winds may also be of relevance, but we caution that the results shown here argue that such winds must be efficient at wholesale ISM removal {\it only} during star formation events that create a prominent bulge (as disk galaxies keep forming stars with enthusiasm).  

\subsection{Quiescent disks?}

\begin{figure}[t]
\begin{center}
\includegraphics[width=8.0cm]{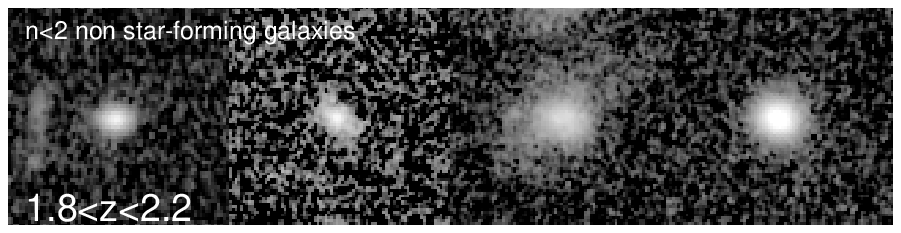}
\includegraphics[width=8.0cm]{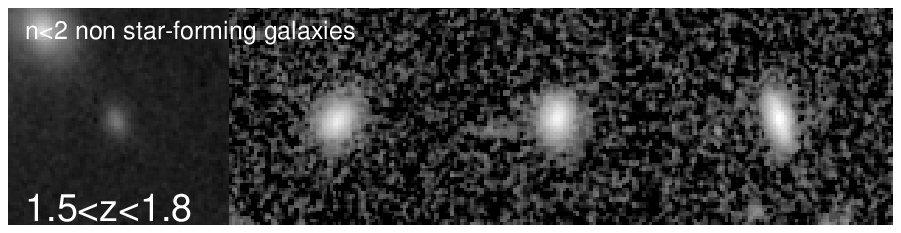}
\caption{F160W Postage stamps of quiescent galaxies with $n < 2$ and stellar surface densities between $10^9$ and $10^{10} M_{\Sun}\,{\rm kpc^{-2}}$.  These galaxies should be star-forming based on their S\'ersic indices, but they are in fact quiescent.  The postage stamps, within each class, are ordered by stellar mass (ordered left to right).  At all redshifts, all postage stamps are 40 physical kpc on a side, and are scaled to a constant 'stellar mass density' (total intensity is scaled to total stellar mass, meaning that if the stellar M/L is constant over the face of the galaxy this postage stamp should reflect the stellar mass density) and are displayed using asinh scaling (linear at low intensity, and logarithmic at higher surface brightness; \protect\citealp{luptitudes}).  
\label{fig:nsfstmp} 
}
\end{center}
\end{figure}

There has been much recent discussion of the presence and importance of stellar disks in quiescent galaxies.  At $z\ga 1.5$, \citet{stockton04}, \citet{mcgrath08}, \citet{vd08} and \citet{vdw11} have argued that most quiescent galaxies with masses in excess of $10^{11} M_{\sun}$ have prominent stellar disks.  \citet{bundy10} have explored this issue at $z < 1.2$, finding that a large fraction of quiescent galaxies have disks in addition to significant bulge components; indeed, the existence of S0s and disky ellipticals in all environments is well-known \citep[see, e.g.,][and references therein]{vandenbergh09}.  The highly-inclined fraction of these systems are clearly visible in Fig.\ \ref{fig:colser}, as the elongated symbols with low $b/a$.  Such systems appear to be somewhat more common at $z>1.5$, but are present (especially at lower S\'ersic indices $n \sim 2-3$) at all redshifts.  

There are two comments that we wish to make about quiescent disks.  Firstly, the vast majority of these systems have S\'ersic indices $n \ga 2$.  In \citet{mcgrath08} and \citet{vdw11}, bulge/disk decompositions were carried out, and showed that these systems with relatively high S\'ersic index are also well-explained as composite bulge/disk systems with relatively large bulges.  Furthermore, kinematic studies of local quiescent galaxies have demonstrated that the vast majority of quiescent galaxies have significant rotation \citep{emsellem11}, and the incidence of strong rotation signatures in quiescent galaxies does not change from $z \sim 1$ to the present day (at $\sim 60$\%, \citealp{vdw_vdm08}).  The picture that emerges is that the vast majority of quiescent systems have undergone some event that both steepens their light profile (either by creation of a distinct bulge component, or simply by steepening the light profile) but manages to retain a significant fraction of the system's original angular momentum in preserving a disk component.  In the context of galaxy merging, such systems are a relatively natural outcome of merging between disk galaxies with even modest gas fractions, where higher mass ratio minor mergers (or mergers between more gas-rich systems) lead to progressively more disk-dominated remnants (see, e.g., \citealp{naab06}, \citealp{hopkins09}, \citealp{hof10}).   

Secondly, a small fraction of quiescent systems appear to have little in the way of a bulge component \citep{bell_disk08,stockton04,mcgrath08,bundy10}; the examples in \citet{mcgrath08} have been very carefully documented.  At least a few of the examples in Fig.\ \ref{fig:nsfstmp} appear to have genuinely low-concentration light profiles, and at least one is an inclined disk.  Given that such systems in the local universe appear to all be satellites in high mass (group or cluster) halos \citep{bell_disk08}, it will be interesting to explore the environments of such galaxies, as a function of redshift, and help to elucidate the extent to which disk-only quiescent galaxies are the products of stripping of their cold gas content by hot gas in a deep potential well (in this context, it is worth noting that the $z \sim 1.5$ systems in \citealp{mcgrath08} were chosen to be in fields near radio-loud $z \sim 1.5$ QSOs, and may in fact reside in overdensities).  In this context, we note that R.\ Bassett et al.\ (in preparation) have explored this issue in a narrow redshift slice with the dataset used in this paper, focusing on an overdensity at $z \sim 1.62$, finding weak evidence of an increase in the fraction of quiescent galaxies with low S\'ersic indices near the largest overdensity in the UDS. Their result, although it must be explored with larger samples, appears consistent with the notion that disk-only quiescent galaxies should be best interpreted as being gas-free owing to external (e.g., environmental) mechanisms, not internal processes.  

\subsection{On the nature of high S\'ersic index star-forming galaxies}

\begin{figure}[t]
\begin{center}
\includegraphics[width=8.0cm]{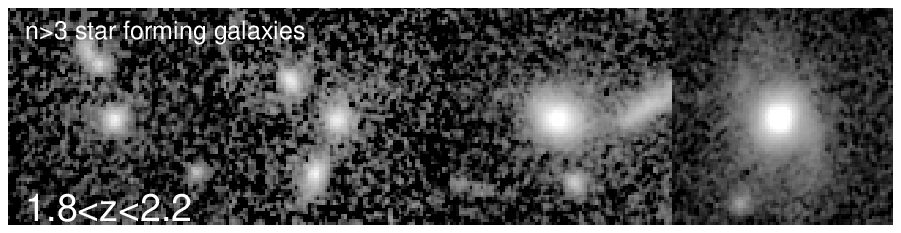}
\includegraphics[width=8.0cm]{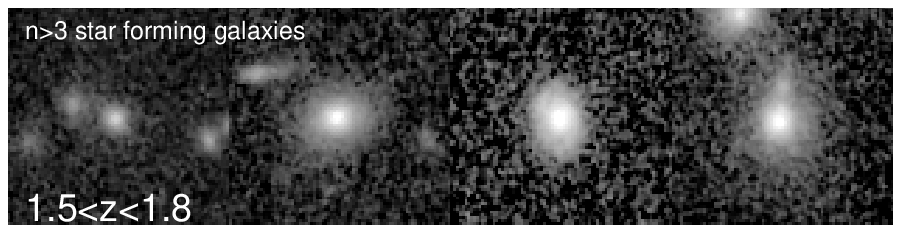}
\includegraphics[width=8.0cm]{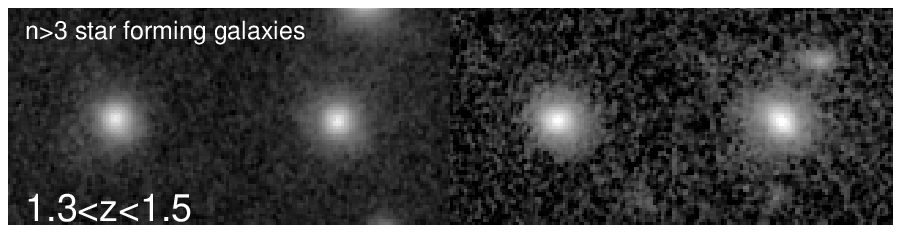}
\includegraphics[width=8.0cm]{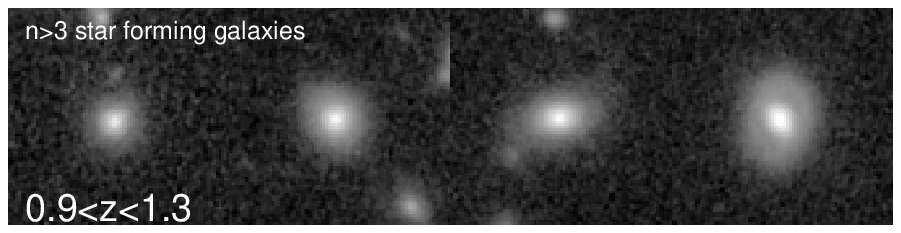}
\includegraphics[width=8.0cm]{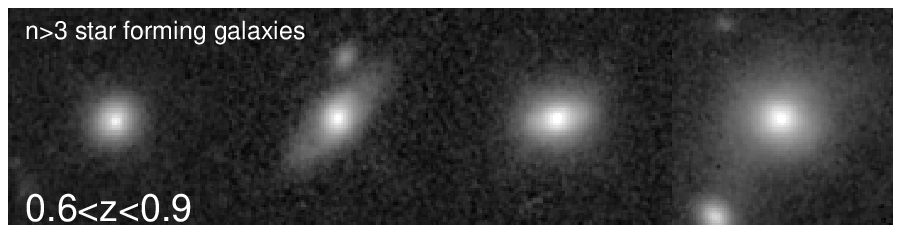}
\caption{F160W Postage stamps of star-forming galaxies with $n > 3$ and stellar surface densities between $10^9$ and $10^{10} M_{\Sun}\,{\rm kpc^{-2}}$.  The postage stamps, within each class, are ordered by stellar mass (ordered left to right).  At all redshifts, all postage stamps are 40 physical kpc on a side, and are scaled to a constant 'stellar mass density' (total intensity is scaled to total stellar mass, meaning that if the stellar M/L is constant over the face of the galaxy this postage stamp should reflect the stellar mass density) and are displayed using asinh scaling (linear at low intensity, and logarithmic at higher surface brightness; \protect\citealp{luptitudes}).  
\label{fig:sfstmp} 
}
\end{center}
\end{figure}

Figs.\ \ref{fig:colser} and \ref{fig:colsep} show a population of star-forming galaxies with high S\'ersic index.  We show examples of such systems in Fig.\ \ref{fig:sfstmp} in 5 different redshift bins.  One can see that these high S\'ersic index, star-forming systems fall broadly into two types of systems.  At lower redshifts, there are some clear examples of disks with very prominent bulges; such galaxies appear to be the star-forming counterparts of the quiescent disks discussed above, and appear to be systems that have either retained, or `re-grown', a substantial disk during/after the bulge formation process \citep{baugh96,kannappan09,benson10}.  At all redshifts, there are what appear to be genuinely spheroidal galaxies, but in many cases with significant asymmetries, and in some cases evidence of interactions (e.g., nearby companions, tidal tails).  It will be interesting to examine such systems in the future to understand the degree of overlap between these systems and `blue spheroids' \citep{menanteau01,boris_thesis,kannappan09,gyory10}, post-starburst galaxies \citep{vergani10}, or ongoing `bulge-building' starbursts \citep{wuyts11} and to ask if the properties of these galaxies are more consistent with an interpretation as a quiescent galaxy in formation, a galaxy in the midst of bulge building through instabilities in the stellar disk and cold inflowing gas \citep{dekel09,ceverino10}, or a galaxy in the early stages of re-growing a disk \citep{kannappan09}.  

\subsection{Future steps}

There are a number of possible future directions for pursuing this line of study further.  A clear improvement path involves the use of more accurate photometric spectroscopic redshifts, which helps improve both the accuracy and reliability of redshift and stellar population estimates, and opens up possibilities to measure low-resolution spectral parameters like the 4000{\AA} break strength \citep[e.g.,][]{whitaker11,kriek11,vd11}.  Further sharpening the analysis would be possible with the addition of reliable bulge/disk decompositions: the hypothesis that the existence of a prominent bulge is necessary but not sufficient to shut off star formation can be better tested, and additional parameters can be explored for possible relevance (e.g., bulge/total ratio, bulge S\'ersic index, or bulge mass; see, e.g., \citealp{drory07} for an exploration of some of these issues with a low-redshift sample, or V.\ Bruce et al.\ in preparation).  More detailed exploration of the outliers at all redshifts (the quiescent lower S\'ersic index systems, or star-forming systems with high S\'ersic index) may help to illuminate the processes that remove (or keep out) cold gas from galaxies.  Finally, increasing the number statistics at $z<1$ (already underway with, e.g., the GEMS, AEGIS or COSMOS datasets; \citealp{rix04}, \citealp{davis07}, \citealp{scoville07}) and at $z>1$ with the full five-field coverage of CANDELS \citep{grogin11,koe11} is likely to prove useful.

\section{Conclusions}

Motivated by a desire to empirically explore the evolutionary factors that lead to a lack of star formation in galaxies, we have explored the structures and star formation activity of the galaxy population using the {\it HST}/WFC3 F160W imaging data from CANDELS in the UDS field.  We used public photometry and photometric redshifts from \citet{williams09}, and determined rest-frame absolute magnitudes and stellar masses using our own stellar population model fits.  We supplement this with public 24{\micron} data from SpUDS, and separate galaxies into quiescent and star-forming using a combination of optical--near-infrared colors and 24{\micron} information.  For structural information, we use parametric fits of a single S\'ersic profile to the WFC3 F160W imaging data.  We then combine these data for $z>0.6$ galaxies with a sample with similar parameters from the SDSS, to create a $z<0.05$ comparison sample.  We then proceed to explore the evolution of the $M_* > 3 \times 10^{10} M_{\sun}$ galaxy population from $z=2.2$ to the present day.  

We first visualize the evolution of the galaxy population over the last 10\,Gyr by normalizing the sample to a `fixed' comoving volume of $10^5$ Mpc$^3$.  In agreement with a large number of other works, we find that the number density of galaxies with stellar mass in excess of $3 \times 10^{10} M_{\sun}$ increases approximately five-fold from $z \sim 2$ to the present day, and that the number density of quiescent galaxies increases yet more rapidly.  Furthermore, examining the properties of the quiescent galaxy population, we find that the vast majority of those quiescent galaxies have high S\'ersic indices (and, at $z \ga 1$, high inferred velocity dispersion and surface density), a sign that they have a prominent bulge component.  The growth of the quiescent galaxy population appears to be intimately linked with the growth of galaxies with prominent bulges.

Given the rapid evolution of the quiescent galaxy population, we proceed to explore the strength of the correlation between quiescence and four galaxy/structural parameters: stellar mass, `velocity dispersion' $M/R \propto \sigma^{\prime\,2}$ with a S\'ersic index-dependent scaling, stellar surface density and S\'ersic index.  At all redshifts $z<2.2$, stellar mass correlates poorly with quiescence.  It is possible, bearing in mind the inferred $\la 0.2$\,dex scatter between stellar mass and dark matter halo mass at $z \sim 0$ \citep{more09}, that the weakness of this correlation indicates that halo mass {\it alone} is not the main determinant of quiescence (but see also \citealp{wake_halo} and \citealp{wake_quench} for arguments that velocity dispersion correlates with halo mass better than stellar mass, at least at high stellar mass).  

At $z \la 0.05$, we find that quiescence correlates much more strongly with S\'ersic index than either velocity dispersion or surface density.  At $z \ge 0.6$, we find that velocity dispersion, surface density and S\'ersic index correlate well with quiescence, where the correlation of S\'ersic index with quiescence appears marginally stronger.  

All correlations have substantial scatter, however.  Many quiescent systems have prominent disks, although the vast majority of quiescent galaxies with disks also have prominent bulges.  A very small fraction of quiescent galaxies appear to be bulgeless disks. In the local Universe they are all satellite galaxies in galaxy groups/clusters; at higher redshifts we did not explore environmental variables for lack of dynamic range in galaxy environments (see R. Bassett et al. in preparation for first steps in this direction using the CANDELS dataset).  Star-forming systems with high $n$, $M$, $\Sigma$ and $\sigma^{\prime}$ are not particularly uncommon; at high redshifts they appear to be genuinely compact with high $n$, and often show asymmetries or signatures of tidal interactions (one may wish to associate these with the possible remnants of gas-rich dissipational galaxy interactions/mergers), and at lower redshifts there is a mix of similar systems and composite bulge-disk star-forming systems (with large bulges).  

At $z<0.05$, \citet{bell_disk08} concluded that a prominent bulge (and by association, a supermassive black hole) was a necessary but not sufficient condition for a galaxy to turn off its own star formation on galaxy-wide scales (all quiescent galaxies in the centers of their own halos had prominent bulges, but not all galaxies with bulges lack star formation).  This observational association is qualitatively consistent with the AGN feedback paradigm (no supermassive black hole, no ability to shut off star formation).  While there is clearly scope for further investigation of the drivers of quiescence, the evidence assembled here appears to be consistent with this proposition to $z<2.2$, a time interval of more than 10Gyr. 

\acknowledgements
We appreciate the constructive and helpful report from the referee, Pieter van Dokkum.
This work is supported by {\it HST} grant GO-12060.
Support for Program number GO-12060 was provided by NASA through a grant from the Space Telescope Science
Institute, which is operated by the Association of Universities for Research in Astronomy, Incorporated, under NASA
contract NAS5-26555.

This publication makes use of the {\it Sloan Digital
Sky Survey} (SDSS).
Funding for the creation and distribution of the SDSS 
Archive has been provided by the Alfred P.\ Sloan Foundation, the 
Participating Institutions, the National Aeronautics and Space Administration,
the National Science Foundation, the US Department of Energy, 
the Japanese Monbukagakusho, and the Max Planck Society.  The SDSS
Web site is \texttt {http://www.sdss.org/}.  The SDSS Participating
Institutions are the University of Chicago, Fermilab, the Institute 
for Advanced Study, the Japan Participation Group, the Johns Hopkins
University, the Max Planck Institut f\"ur Astronomie, the Max
Planck Institut f\"ur Astrophysik, New Mexico State University, 
Princeton University, the United States Naval Observatory, and 
the University of Washington. 
This publication also makes use of data products from the Two Micron All Sky Survey, which is a joint project of the University of Massachusetts and the Infrared Processing and Analysis Center/California Institute of Technology, funded by the National Aeronautics and Space Administration and the National Science Foundation.  This publication also made use of NASA's Astrophysics Data System 
Bibliographic Services.

\appendix

\section{Appendix A: Normalizing the sample to a constant comoving volume} 
\label{ap:norm}

\begin{figure}[t]
\begin{center}
\includegraphics[width=8.0cm]{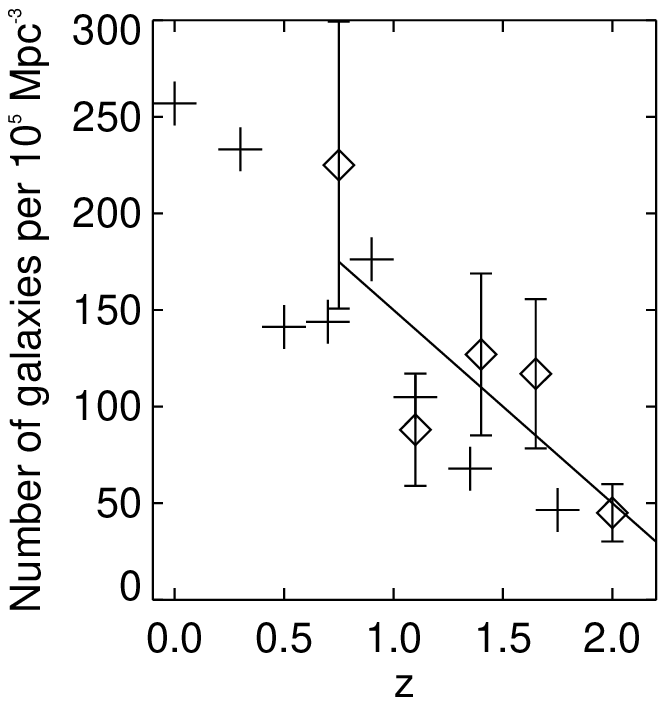}
\caption{The number of galaxies with stellar masses in excess of $3 \times 10^{10} M_{\sun}$ in a volume of $10^5 {\rm Mpc}^3$ measured from the UDS (diamonds, with the dominant sample variance uncertainties as estimated using the method of \citealp{somerville04}; see also \citealp{moster11}), and the number of galaxies with masses in excess of $3 \times 10^{10} M_{\sun}$ (crosses) inferred from the stellar mass functions of \citet[$z>0.2$]{ilbert10} and \citet[local values]{bell03mf}.  The number of galaxies with $M_* > 3 \times 10^{10} M_{\sun}$ in the UDS is consistent to within the significant sample variance and systematic stellar mass uncertainties with those of larger fields.  We choose in this paper to adjust the number of galaxies for display in Fig.\ \protect\ref{fig:colmass} to fit the straight line in this figure, to take out the first-order effects of sample variance on the results (while noting that doing so makes no important difference to our conclusions; the evolution of the population is not subtle). 
\label{fig:ngal} 
}
\end{center}
\end{figure}

In Fig.\ \ref{fig:colmass}, we Monte Carlo subsampled the galaxy population of the UDS and the SDSS to an equivalent comoving volume of $10^5 {\rm Mpc}^3$.  There are two ways to do this: simply rescaling the sample by $10^5 {\rm Mpc}^3 / V(z)$, where $V(z)$ is the comoving volume of that redshift bin (leaving one susceptible to the first-order effects of sample variance), or by rescaling the sample in a given redshift bin to have the number of galaxies expected at that redshift above that mass limit using mass functions derived from much larger surveys (canceling out number density variation from sample variance but leaving behind any systematic variation in galaxy properties that are a function of the variation in the average environment in this light cone as a function of redshift).  

The two approaches are compared in Fig.\ \ref{fig:ngal}.  Diamonds with error bars denote the number of galaxies with stellar masses in excess of $3 \times 10^{10} M_{\sun}$ in the UDS, with error bars denoting the expected degree of sample variance in that bin following the method of \citet{somerville04}.  Crosses at $z>0.2$ show the number of galaxies with stellar mass in excess of $3 \times 10^{10} M_{\sun}$ expected from the stellar mass functions of \citet{ilbert10}.  The cross at low redshift shows the number of galaxies with $M_* > 3 \times 10^{10} M_{\sun}$ from the mass function of \citet{bell03mf}.  The number of galaxies observed in the UDS is broadly consistent with, or perhaps somewhat larger than, the number of galaxies expected from larger surveys, given the substantial sample variance uncertainties.  There are systematic uncertainties also on the crosses; choosing different star formation histories for constructing stellar masses, stellar population models, etc., can give more than a factor of two variation in stellar masses that translates into around 50\% in number density uncertainty. For Fig.\ \ref{fig:colmass}, we chose to rescale the number of galaxies to the smoothly-varying number of galaxies given by the line shown in Fig.\ \ref{fig:ngal} (approximately corresponding to scaling the number of galaxies to larger cosmological surveys).  

None of the results shown in Fig.\ \ref{fig:colmass} depend in any important way on the choice of this scaling method; the evolution of the galaxy population from $z \sim 2$ to the present day is not subtle and is robust to even the significant systematic uncertainties inherent to mass function analyses.  While this scaling in the number of galaxies largely counteracts the worst effects of sample variance, if the properties of galaxies depend strongly on environment there will be a second-order difference between the properties of galaxies sampled from a globally-underdense vs.\ a globally-overdense volume.  In the absence of enough environmental dynamic range to robustly measure its effect within CANDELS at the current time (see \citealp{papovich11} and R.\ Bassett et al., in prep., for a discussion of SFR, galaxy size and structure trends as a function of environment using a known $z=1.62$ galaxy cluster in the CANDELS UDS coverage), we made no attempt to correct for this second-order effect in this paper.


\begin{thebibliography}{133}
\expandafter\ifx\csname natexlab\endcsname\relax\def\natexlab#1{#1}\fi

\bibitem[{{Abazajian} {et~al.}(2004){Abazajian}, { }, { }, \& {et al.}}]{dr2}
{Abazajian}, K., {et al.} 2004, \aj, 128, 502

\bibitem[{{Arnouts} {et~al.}(2007){Arnouts}, {Walcher}, {Le F{\`e}vre},
  {Zamorani}, {Ilbert}, {Le Brun}, {Pozzetti}, {Bardelli}, {Tresse}, {Zucca},
  {Charlot}, {Lamareille}, {McCracken}, {Bolzonella}, {Iovino}, {Lonsdale},
  {Polletta}, {Surace}, {Bottini}, {Garilli}, {Maccagni}, {Picat},
  {Scaramella}, {Scodeggio}, {Vettolani}, {Zanichelli}, {Adami}, {Cappi},
  {Ciliegi}, {Contini}, {de la Torre}, {Foucaud}, {Franzetti}, {Gavignaud},
  {Guzzo}, {Marano}, {Marinoni}, {Mazure}, {Meneux}, {Merighi}, {Paltani},
  {Pell{\`o}}, {Pollo}, {Radovich}, {Temporin}, \& {Vergani}}]{arnouts07}
{Arnouts}, S., et al.\ 2007, \aap, 476, 137

\bibitem[Barden et al.(2012)]{galapagos}
 Barden, M., H\"au{\ss}ler, B., Peng, C.\ Y., McIntosh, D.\ H.\ \& Guo, Y. 2012, \mnras, in press (arXiv:1203.1831)

\bibitem[{{Barnes} \& {Hernquist}(1992)}]{barnes92}
{Barnes}, J.~E., \& {Hernquist}, L. 1992, \araa, 30, 705

\bibitem[{{Baugh} {et~al.}(1996){Baugh}, {Cole}, \& {Frenk}}]{baugh96}
{Baugh}, C.~M., {Cole}, S., \& {Frenk}, C.~S. 1996, \mnras, 283, 1361

\bibitem[{{Bell}(2008)}]{bell_disk08}
{Bell}, E.~F. 2008, \apj, 682, 355

\bibitem[{{Bell} \& {de Jong}(2001)}]{bdj}
{Bell}, E.~F., \& {de Jong}, R.~S. 2001, \apj, 550, 212

\bibitem[{{Bell} {et~al.}(2003){Bell}, {McIntosh}, {Katz}, \&
  {Weinberg}}]{bell03mf}
{Bell}, E.~F., {McIntosh}, D.~H., {Katz}, N., \& {Weinberg}, M.~D. 2003, \apjs,
  149, 289

\bibitem[{{Bell} {et~al.}(2004){Bell}, {Wolf}, {Meisenheimer}, {Rix}, {Borch},
  {Dye}, {Kleinheinrich}, {Wisotzki}, \& {McIntosh}}]{bell04c17}
{Bell}, E.~F., et al.\ 2004,
  \apj, 608, 752

\bibitem[{{Bell} {et~al.}(2007){Bell}, {Zheng}, {Papovich}, {Borch}, {Wolf}, \&
  {Meisenheimer}}]{bellcont}
{Bell}, E.~F., {Zheng}, X.~Z., {Papovich}, C., {Borch}, A., {Wolf}, C., \&
  {Meisenheimer}, K. 2007, \apj, 663, 834

\bibitem[{{Benson} {et~al.}(2003){Benson}, {Bower}, {Frenk}, {Lacey}, {Baugh},
  \& {Cole}}]{benson03}
{Benson}, A.~J., {Bower}, R.~G., {Frenk}, C.~S., {Lacey}, C.~G., {Baugh},
  C.~M., \& {Cole}, S. 2003, \apj, 599, 38

\bibitem[{{Benson} \& {Devereux}(2010)}]{benson10}
{Benson}, A.~J., \& {Devereux}, N. 2010, \mnras, 402, 2321

\bibitem[Bertin et al.(2002)]{bertin02} Bertin, G., Ciotti, L.\ \& Del Principe, M. 2002, \aap, 386, 149

\bibitem[{{Best} {et~al.}(2006){Best}, {Kaiser}, {Heckman}, \&
  {Kauffmann}}]{best06}
{Best}, P.~N., {Kaiser}, C.~R., {Heckman}, T.~M., \& {Kauffmann}, G. 2006,
  \mnras, 368, L67

\bibitem[Bezanson et al.(2011)]{bezanson11} Bezanson, R., van Dokkum, P.\ G., Franx, M., Brammer, G.\ B., et al. 2011, \apj, 737, L31

\bibitem[{{Birnboim} {et~al.}(2007){Birnboim}, {Dekel}, \&
  {Neistein}}]{birnboim07}
{Birnboim}, Y., {Dekel}, A., \& {Neistein}, E. 2007, \mnras, 380, 339

\bibitem[{{Blanton} {et~al.}(2005){Blanton}, { }, { }, \& {et al.}}]{vagc}
{Blanton}, M.~R., {et al.} 2005, \aj, 129, 2562

\bibitem[{{Blanton} {et~al.}(2003){Blanton}, {Hogg}, {Bahcall}, \& {et
  al.}}]{blanton03prop}
{Blanton}, M.~R., {Hogg}, D.~W., {Bahcall}, N.~A., \& {et al.} 2003, \apj, 594,
  186

\bibitem[{{Borch} {et~al.}(2006){Borch}, {Meisenheimer}, {Bell}, {Rix}, {Wolf},
  {Dye}, {Kleinheinrich}, {Kovacs}, \& {Wisotzki}}]{borch06}
{Borch}, A., et al.\ 2006, \aap,
  453, 869

\bibitem[{{Brammer} {et~al.}(2008){Brammer}, {van Dokkum}, \& {Coppi}}]{eazy}
{Brammer}, G.~B., {van Dokkum}, P.~G., \& {Coppi}, P. 2008, \apj, 686, 1503

\bibitem[{{Brammer} {et~al.}(2011){Brammer}, {Whitaker}, {van Dokkum},
  {Marchesini}, {Franx}, {Kriek}, {Labbe}, {Lee}, {Muzzin}, {Quadri},
  {Rudnick}, \& {Williams}}]{brammer11}
{Brammer}, G.~B., et al.\ 2011, \apj, 739, 24

\bibitem[{{Brinchmann} {et~al.}(2004){Brinchmann}, {Charlot}, {White},
  {Tremonti}, {Kauffmann}, {Heckman}, \& {Brinkmann}}]{brinchmann04}
{Brinchmann}, J., {Charlot}, S., {White}, S.~D.~M., {Tremonti}, C.,
  {Kauffmann}, G., {Heckman}, T., \& {Brinkmann}, J. 2004, \mnras, 351, 1151

\bibitem[{{Brown} {et~al.}(2007){Brown}, {Dey}, {Jannuzi}, {Brand}, {Benson},
  {Brodwin}, {Croton}, \& {Eisenhardt}}]{brown07}
{Brown}, M.~J.~I., {Dey}, A., {Jannuzi}, B.~T., {Brand}, K., {Benson}, A.~J.,
  {Brodwin}, M., {Croton}, D.~J., \& {Eisenhardt}, P.~R. 2007, \apj, 654, 858

\bibitem[{{Bruzual} \& {Charlot}(2003)}]{bc03}
{Bruzual}, G., \& {Charlot}, S. 2003, \mnras, 344, 1000

\bibitem[{Bundy} {et~al.}(2005)]{bundy05}
{Bundy}, K., Ellis, R.\ S., \& Conselice, C.\ J. 2005, \apj, 625, 621

\bibitem[{{Bundy} {et~al.}(2010){Bundy}, {Scarlata}, {Carollo}, {Ellis},
  {Drory}, {Hopkins}, {Salvato}, {Leauthaud}, {Koekemoer}, {Murray}, {Ilbert},
  {Oesch}, {Ma}, {Capak}, {Pozzetti}, \& {Scoville}}]{bundy10}
{Bundy}, K., et al.\ 2010, \apj, 719, 1969

\bibitem[{{Calzetti}(2001)}]{calzetti01}
{Calzetti}, D. 2001, \pasp, 113, 1449

\bibitem[{{Cameron} {et~al.}(2010){Cameron}, {Carollo}, {Oesch}, {Bouwens},
  {Illingworth}, {Trenti}, {Labbe}, \& {Magee}}]{cameron10}
{Cameron}, E., {Carollo}, C.~M., {Oesch}, P.~A., {Bouwens}, R.~J.,
  {Illingworth}, G.~D., {Trenti}, M., {Labbe}, I., \& {Magee}, D. 2011, \apj, 743, 146

\bibitem[{{Cassata} {et~al.}(2011){Cassata}, {Giavalisco}, {Guo}, {Renzini},
  {Ferguson}, {Koekemoer}, {Salimbeni}, {Scarlata}, {Grogin}, {Conselice},
  {Dahlen}, {Lotz}, {Dickinson}, \& {Lin}}]{cassata11}
{Cassata}, P., et al.\ 2011, \apj, 743, 96

\bibitem[{{Cattaneo} {et~al.}(2006){Cattaneo}, {Dekel}, {Devriendt},
  {Guiderdoni}, \& {Blaizot}}]{cattaneo06}
{Cattaneo}, A., {Dekel}, A., {Devriendt}, J., {Guiderdoni}, B., \& {Blaizot},
  J. 2006, \mnras, 370, 1651

\bibitem[Ceverino, Dekel \& Bournaud(2010)]{ceverino10} Ceverino, D., Dekel, A., \& Bournaud, F.\ 2010, \mnras, 404, 2151

\bibitem[{{Chabrier}(2003)}]{chabrier}
{Chabrier}, G. 2003, \pasp, 115, 763

\bibitem[Cheung et al.(2012)]{cheung}
Cheung, E., Faber, S.\ M., Koo, D.\ C., Dutton, A.\ A., et al. 2012, submitted to \apj

\bibitem[{{Croton} {et~al.}(2006){Croton}, {Springel}, {White}, {De Lucia},
  {Frenk}, {Gao}, {Jenkins}, {Kauffmann}, {Navarro}, \& {Yoshida}}]{croton06}
{Croton}, D.~J., et al.\ 2006, \mnras, 365, 11

\bibitem[{{Dav{\'e}} {et~al.}(2011){Dav{\'e}}, {Oppenheimer}, \&
  {Finlator}}]{dave11}
{Dav{\'e}}, R., {Oppenheimer}, B.~D., \& {Finlator}, K. 2011, \mnras, 867

\bibitem[{{Davis} {et~al.}(2007){Davis}, {Guhathakurta}, {Konidaris}, {Newman},
  {Ashby}, {Biggs}, {Barmby}, {Bundy}, {Chapman}, {Coil}, {Conselice},
  {Cooper}, {Croton}, {Eisenhardt}, {Ellis}, {Faber}, {Fang}, {Fazio},
  {Georgakakis}, {Gerke}, {Goss}, {Gwyn}, {Harker}, {Hopkins}, {Huang},
  {Ivison}, {Kassin}, {Kirby}, {Koekemoer}, {Koo}, {Laird}, {Le Floc'h}, {Lin},
  {Lotz}, {Marshall}, {Martin}, {Metevier}, {Moustakas}, {Nandra}, {Noeske},
  {Papovich}, {Phillips}, {Rich}, {Rieke}, {Rigopoulou}, {Salim},
  {Schiminovich}, {Simard}, {Smail}, {Small}, {Weiner}, {Willmer}, {Willner},
  {Wilson}, {Wright}, \& {Yan}}]{davis07}
{Davis}, M., et al.\ 2007, \apjl, 660, L1

\bibitem[{{Dekel} \& {Birnboim}(2006)}]{dekel06}
{Dekel}, A., \& {Birnboim}, Y. 2006, \mnras, 368, 2

\bibitem[{{Dekel} \& {Birnboim}(2008)}]{db07}
---. 2008, \mnras, 383, 119

\bibitem[Dekel, Sari \& Ceverino(2009)]{dekel09}
  Dekel, A., Sari, R.\ \& Ceverino, D.\ 2009, \apj, 703, 785

\bibitem[{{Dom{\'{\i}}nguez S{\'a}nchez} {et~al.}(2011){Dom{\'{\i}}nguez
  S{\'a}nchez}, {Pozzi}, {Gruppioni}, {Cimatti}, {Ilbert}, {Pozzetti},
  {McCracken}, {Capak}, {Le Floch}, {Salvato}, {Zamorani}, {Carollo},
  {Contini}, {Kneib}, {Le F{\'e}vre}, {Lilly}, {Mainieri}, {Renzini},
  {Scodeggio}, {Bardelli}, {Bolzonella}, {Bongiorno}, {Caputi}, {Coppa},
  {Cucciati}, {de la Torre}, {de Ravel}, {Franzetti}, {Garilli}, {Iovino},
  {Kampczyk}, {Knobel}, {Kovac}, {Lamareille}, {Le Borgne}, {Le Brun}, {Maier},
  {Mignoli}, {Pell{\'o}}, {Peng}, {P{\'e}rez-Montero}, {Ricciardelli},
  {Silverman}, {Tanaka}, {Tasca}, {Tresse}, {Vergani}, \& {Zucca}}]{dom11}
{Dom{\'{\i}}nguez S{\'a}nchez}, H., et al.\ 2011, \mnras, 417, 900

\bibitem[{{Donley} {et~al.}(2008){Donley}, {Rieke}, {P{\'e}rez-Gonz{\'a}lez},
  \& {Barro}}]{donley08}
{Donley}, J.~L., {Rieke}, G.~H., {P{\'e}rez-Gonz{\'a}lez}, P.~G., \& {Barro},
  G. 2008, \apj, 687, 111

\bibitem[{{Drory} \& {Fisher}(2007)}]{drory07}
{Drory}, N., \& {Fisher}, D.~B. 2007, \apj, 664, 640

\bibitem[{{Dunne} {et~al.}(2009){Dunne}, {Ivison}, {Maddox}, {Cirasuolo},
  {Mortier}, {Foucaud}, {Ibar}, {Almaini}, {Simpson}, \& {McLure}}]{dunne09}
{Dunne}, L., et al.\
  2009, \mnras, 394, 3

\bibitem[{{Elbaz} {et~al.}(2011){Elbaz}, {Dickinson}, {Hwang}, {Diaz-Santos},
  {Magdis}, {Magnelli}, {Le Borgne}, {Galliano}, {Pannella}, {Chanial},
  {Armus}, {Charmandaris}, {Daddi}, {Aussel}, {Popesso}, {Kartaltepe},
  {Altieri}, {Valtchanov}, {Coia}, {Dannerbauer}, {Dasyra}, {Leiton},
  {Mazzarella}, {Buat}, {Burgarella}, {Chary}, {Gilli}, {Ivison}, {Juneau},
  {LeFloc'h}, {Lutz}, {Morrison}, {Mullaney}, {Murphy}, {Pope}, {Scott},
  {Alexander}, {Brodwin}, {Calzetti}, {Cesarsky}, {Charlot}, {Dole},
  {Eisenhardt}, {Ferguson}, {Foerster-Schreiber}, {Frayer}, {Giavalisco},
  {Huynh}, {Koekemoer}, {Papovich}, {Reddy}, {Surace}, {Teplitz}, {Yun}, \&
  {Wilson}}]{elbaz11}
{Elbaz}, D., et al.\ 2011, \aap, 533, 119

\bibitem[{{Emsellem} {et~al.}(2011){Emsellem}, {Cappellari}, {Krajnovi{\'c}},
  {Alatalo}, {Blitz}, {Bois}, {Bournaud}, {Bureau}, {Davies}, {Davis}, {de
  Zeeuw}, {Khochfar}, {Kuntschner}, {Lablanche}, {McDermid}, {Morganti},
  {Naab}, {Oosterloo}, {Sarzi}, {Scott}, {Serra}, {van de Ven}, {Weijmans}, \&
  {Young}}]{emsellem11}
{Emsellem}, E., et al.\ 2011, \mnras, 414, 888

\bibitem[{{Faber} {et~al.}(2007){Faber}, {Willmer}, {Wolf}, {Koo}, {Weiner},
  {Newman}, {Im}, {Coil}, {Conroy}, {Cooper}, {Davis}, {Finkbeiner}, {Gerke},
  {Gebhardt}, {Groth}, {Guhathakurta}, {Harker}, {Kaiser}, {Kassin},
  {Kleinheinrich}, {Konidaris}, {Kron}, {Lin}, {Luppino}, {Madgwick},
  {Meisenheimer}, {Noeske}, {Phillips}, {Sarajedini}, {Schiavon}, {Simard},
  {Szalay}, {Vogt}, \& {Yan}}]{faber07}
{Faber}, S.~M., et al.\ 2007, \apj, 665, 265

\bibitem[{{Fabian} {et~al.}(2006){Fabian}, {Sanders}, {Taylor}, {Allen},
  {Crawford}, {Johnstone}, \& {Iwasawa}}]{fabian06}
{Fabian}, A.~C., {Sanders}, J.~S., {Taylor}, G.~B., {Allen}, S.~W., {Crawford},
  C.~S., {Johnstone}, R.~M., \& {Iwasawa}, K. 2006, \mnras, 366, 417

\bibitem[{{Fioc} \& {Rocca-Volmerange}(1997)}]{pegase}
{Fioc}, M., \& {Rocca-Volmerange}, B. 1997, \aap, 326, 950

\bibitem[{{Fontana} {et~al.}(2009){Fontana}, {Santini}, {Grazian},
  {Pentericci}, {Fiore}, {Castellano}, {Giallongo}, {Menci}, {Salimbeni},
  {Cristiani}, {Nonino}, \& {Vanzella}}]{fontana09}
{Fontana}, A., et al.\ 2009, \aap, 501, 15

\bibitem[{{Franx} {et~al.}(2008){Franx}, {van Dokkum}, {Schreiber}, {Wuyts},
  {Labb{\'e}}, \& {Toft}}]{franx08}
{Franx}, M., {van Dokkum}, P.~G., {Schreiber}, N.~M.~F., {Wuyts}, S.,
  {Labb{\'e}}, I., \& {Toft}, S. 2008, \apj, 688, 770

\bibitem[{{Furusawa} {et~al.}(2008){Furusawa}, {Kosugi}, {Akiyama}, {Takata},
  {Sekiguchi}, {Tanaka}, {Iwata}, {Kajisawa}, {Yasuda}, {Doi}, {Ouchi},
  {Simpson}, {Shimasaku}, {Yamada}, {Furusawa}, {Morokuma}, {Ishida}, {Aoki},
  {Fuse}, {Imanishi}, {Iye}, {Karoji}, {Kobayashi}, {Kodama}, {Komiyama},
  {Maeda}, {Miyazaki}, {Mizumoto}, {Nakata}, {Noumaru}, {Ogasawara}, {Okamura},
  {Saito}, {Sasaki}, {Ueda}, \& {Yoshida}}]{furusawa08}
{Furusawa}, H., et al.\ 2008,
  \apjs, 176, 1

\bibitem[Gallazzi et al.(2005)]{gallazzi05}
  Gallazzi, A., Charlot, S., Brinchmann, J., White, S.\ D.\ M.\ \& Tremonti, C.\ A. 2005, \mnras, 362, 41

\bibitem[Graham \& Driver(2007)]{graham07}
  Graham, A.\ W.\ \& Driver, S.\ P.\ 2007, \apj, 655, 77

\bibitem[{{Grogin} {et~al.}(2011){Grogin}, {Kocevski}, {Faber}, {Ferguson},
  {Koekemoer}, {Riess}, {Acquaviva}, {Alexander}, {Almaini}, {Ashby}, {Barden},
  {Bell}, {Bournaud}, {Brown}, {Caputi}, {Casertano}, {Cassata}, {Challis},
  {Chary}, {Cheung}, {Cirasuolo}, {Conselice}, {Roshan Cooray}, {Croton},
  {Daddi}, {Dahlen}, {Dav{\'e}}, {de Mello}, {Dekel}, {Dickinson}, {Dolch},
  {Donley}, {Dunlop}, {Dutton}, {Elbaz}, {Fazio}, {Filippenko}, {Finkelstein},
  {Fontana}, {Gardner}, {Garnavich}, {Gawiser}, {Giavalisco}, {Grazian}, {Guo},
  {Hathi}, {H{\"a}ussler}, {Hopkins}, {Huang}, {Huang}, {Jha}, {Kartaltepe},
  {Kirshner}, {Koo}, {Lai}, {Lee}, {Li}, {Lotz}, {Lucas}, {Madau}, {McCarthy},
  {McGrath}, {McIntosh}, {McLure}, {Mobasher}, {Moustakas}, {Mozena}, {Nandra},
  {Newman}, {Niemi}, {Noeske}, {Papovich}, {Pentericci}, {Pope}, {Primack},
  {Rajan}, {Ravindranath}, {Reddy}, {Renzini}, {Rix}, {Robaina}, {Rodney},
  {Rosario}, {Rosati}, {Salimbeni}, {Scarlata}, {Siana}, {Simard}, {Smidt},
  {Somerville}, {Spinrad}, {Straughn}, {Strolger}, {Telford}, {Teplitz},
  {Trump}, {van der Wel}, {Villforth}, {Wechsler}, {Weiner}, {Wiklind}, {Wild},
  {Wilson}, {Wuyts}, {Yan}, \& {Yun}}]{grogin11}
{Grogin}, et al.\ 2011, \apjs, 197, 35

\bibitem[{{G{\"u}ltekin} {et~al.}(2009){G{\"u}ltekin}, {Richstone}, {Gebhardt},
  {Lauer}, {Tremaine}, {Aller}, {Bender}, {Dressler}, {Faber}, {Filippenko},
  {Green}, {Ho}, {Kormendy}, {Magorrian}, {Pinkney}, \& {Siopis}}]{gueltekin09}
{G{\"u}ltekin}, K., et al.\ 2009, \apj, 698, 198

\bibitem[{{Guo} \& {Oh}(2008)}]{guo07}
{Guo}, F., \& {Oh}, S.~P. 2008, \mnras, 384, 251

\bibitem[{{Guo} {et~al.}(2009){Guo}, {McIntosh}, {Mo}, {Katz}, {van den Bosch},
  {Weinberg}, {Weinmann}, {Pasquali}, \& {Yang}}]{guo09}
{Guo}, Y., et al.\ 2009,
  \mnras, 398, 1129

\bibitem[{{Gy{\H o}ry} \& {Bell}(2010)}]{gyory10}
{Gy{\H o}ry}, Z., \& {Bell}, E.~F. 2010, \apj, 724, 694

\bibitem[{{H{\"a}ring} \& {Rix}(2004)}]{haering04}
{H{\"a}ring}, N., \& {Rix}, H.-W. 2004, \apjl, 604, L89

\bibitem[{{H{\"a}u{\ss}ler}(2007)}]{boris_thesis}
{H{\"a}u{\ss}ler}, B. 2007, PhD thesis, Max-Planck-Institut f{\"u}r Astronomie,
  Heidelberg, Germany

\bibitem[{{H{\"a}ussler} {et~al.}(2007){H{\"a}ussler}, {McIntosh}, {Barden},
  {Bell}, {Rix}, {Borch}, {Beckwith}, {Caldwell}, {Heymans}, {Jahnke}, {Jogee},
  {Koposov}, {Meisenheimer}, {S{\'a}nchez}, {Somerville}, {Wisotzki}, \&
  {Wolf}}]{boris07}
{H{\"a}ussler}, B., et al.\
  2007, \apjs, 172, 615

\bibitem[{{Hernquist}(1993)}]{hernquist93}
{Hernquist}, L. 1993, \apj, 409, 548

\bibitem[{{Hoffman} {et~al.}(2010){Hoffman}, {Cox}, {Dutta}, \&
  {Hernquist}}]{hof10}
{Hoffman}, L., {Cox}, T.~J., {Dutta}, S., \& {Hernquist}, L. 2010, \apj, 723,
  818

\bibitem[{{Hopkins} {et~al.}(2010){Hopkins}, {Bundy}, {Croton}, {Hernquist},
  {Keres}, {Khochfar}, {Stewart}, {Wetzel}, \& {Younger}}]{hopkins10}
{Hopkins}, P.~F., et al.\ 2010, \apj,
  715, 202

\bibitem[{{Hopkins} {et~al.}(2009){Hopkins}, {Cox}, {Dutta}, {Hernquist},
  {Kormendy}, \& {Lauer}}]{hopkins09}
{Hopkins}, P.~F., {Cox}, T.~J., {Dutta}, S.~N., {Hernquist}, L., {Kormendy},
  J., \& {Lauer}, T.~R. 2009, \apjs, 181, 135

\bibitem[{{Ilbert} {et~al.}(2010){Ilbert}, {Salvato}, {Le Floc'h}, {Aussel},
  {Capak}, {McCracken}, {Mobasher}, {Kartaltepe}, {Scoville}, {Sanders},
  {Arnouts}, {Bundy}, {Cassata}, {Kneib}, {Koekemoer}, {Le F{\`e}vre}, {Lilly},
  {Surace}, {Taniguchi}, {Tasca}, {Thompson}, {Tresse}, {Zamojski}, {Zamorani},
  \& {Zucca}}]{ilbert10}
{Ilbert}, O., et al.\ 2010, \apj, 709, 644

\bibitem[{{Johansson} {et~al.}(2009){Johansson}, {Naab}, \&
  {Ostriker}}]{johansson09grav}
{Johansson}, P.~H., {Naab}, T., \& {Ostriker}, J.~P. 2009, \apjl, 697, L38

\bibitem[{{Kannappan} {et~al.}(2009){Kannappan}, {Guie}, \&
  {Baker}}]{kannappan09}
{Kannappan}, S.~J., {Guie}, J.~M., \& {Baker}, A.~J. 2009, \aj, 138, 579

\bibitem[{{Karim} {et~al.}(2011){Karim}, {Schinnerer},
  {Mart{\'{\i}}nez-Sansigre}, {Sargent}, {van der Wel}, {Rix}, {Ilbert},
  {Smol{\v c}i{\'c}}, {Carilli}, {Pannella}, {Koekemoer}, {Bell}, \&
  {Salvato}}]{karim11}
{Karim}, A., et al.\ 2011, \apj, 730, 61

\bibitem[{{Kartaltepe} {et~al.}(2010){Kartaltepe}, {Sanders}, {Le Floc'h},
  {Frayer}, {Aussel}, {Arnouts}, {Ilbert}, {Salvato}, {Scoville}, {Surace},
  {Yan}, {Brusa}, {Capak}, {Caputi}, {Carollo}, {Civano}, {Elvis}, {Faure},
  {Hasinger}, {Koekemoer}, {Lee}, {Lilly}, {Liu}, {McCracken}, {Schinnerer},
  {Smol{\v c}i{\'c}}, {Taniguchi}, {Thompson}, \& {Trump}}]{kartaltepe10}
{Kartaltepe}, J.~S., et al.\ 2010, \apj, 709, 572

\bibitem[{{Kauffmann} \& {Haehnelt}(2000)}]{kauf00}
{Kauffmann}, G., \& {Haehnelt}, M. 2000, \mnras, 311, 576

\bibitem[{{Kauffmann} {et~al.}(2003){Kauffmann}, {Heckman}, {White}, {Charlot},
  {Tremonti}, {Peng}, {Seibert}, {Brinkmann}, {Nichol}, {SubbaRao}, \&
  {York}}]{kauf03_dens}
{Kauffmann}, G., et al.\ 2003, \mnras, 341, 54

\bibitem[{{Kere{\v s}} {et~al.}(2005){Kere{\v s}}, {Katz}, {Weinberg}, \&
  {Dav{\'e}}}]{keres05}
{Kere{\v s}}, D., {Katz}, N., {Weinberg}, D.~H., \& {Dav{\'e}}, R. 2005,
  \mnras, 363, 2

\bibitem[{{Khochfar} \& {Ostriker}(2008)}]{khochfar07}
{Khochfar}, S., \& {Ostriker}, J.~P. 2008, \apj, 680, 54

\bibitem[{{Koekemoer} {et~al.}(2011){Koekemoer}, {Faber}, {Ferguson}, {Grogin},
  {Kocevski}, {Koo}, {Lai}, {Lotz}, {Lucas}, {McGrath}, {Ogaz}, {Rajan},
  {Riess}, {Rodney}, {Strolger}, {Casertano}, {Dahlen}, {Dickinson}, {Dolch},
  {Fontana}, {Giavalisco}, {Grazian}, {Guo}, {Hathi}, {Huang}, {van der Wel},
  {Yan}, {Acquaviva}, {Almaini}, {Ashby}, {Barden}, {Bell}, {Bournaud},
  {Brown}, {Caputi}, {Cassata}, {Challis}, {Chary}, {Cheung}, {Cirasuolo},
  {Conselice}, {Roshan Cooray}, {Croton}, {Daddi}, {Dav{\'e}}, {de Mello}, {de
  Ravel}, {Dekel}, {Donley}, {Dunlop}, {Dutton}, {Elbaz}, {Fazio},
  {Filippenko}, {Finkelstein}, {Frazer}, {Gardner}, {Garnavich}, {Gawiser},
  {Gruetzbauch}, {Hartley}, {H{\"a}ussler}, {Herrington}, {Hopkins}, {Huang},
  {Jha}, {Johnson}, {Kartaltepe}, {Khostovan}, {Kirshner}, {Lani}, {Lee}, {Li},
  {Madau}, {McCarthy}, {McIntosh}, {McLure}, {McPartland}, {Mobasher},
  {Moreira}, {Mortlock}, {Moustakas}, {Mozena}, {Nandra}, {Newman}, {Nielsen},
  {Niemi}, {Noeske}, {Papovich}, {Pentericci}, {Pope}, {Primack},
  {Ravindranath}, {Reddy}, {Renzini}, {Rix}, {Robaina}, {Rosario}, {Rosati},
  {Salimbeni}, {Scarlata}, {Siana}, {Simard}, {Smidt}, {Snyder}, {Somerville},
  {Spinrad}, {Straughn}, {Telford}, {Teplitz}, {Trump}, {Vargas}, {Villforth},
  {Wagner}, {Wandro}, {Wechsler}, {Weiner}, {Wiklind}, {Wild}, {Wilson},
  {Wuyts}, \& {Yun}}]{koe11}
{Koekemoer}, A.~M., et al.\ 2011, \apjs, 197, 36

\bibitem[{Kriek} {et~al.}(2011)]{kriek11}
{Kriek}, M., van Dokkum, P.\ G., Whitaker, K.\ E., Labb\'e, I., Franx, M., \& Brammer, G.\ B., 2011, \apj, 743, 168

\bibitem[{{Kriek} {et~al.}(2010){Kriek}, {Labb{\'e}}, {Conroy}, {Whitaker},
  {van Dokkum}, {Brammer}, {Franx}, {Illingworth}, {Marchesini}, {Muzzin},
  {Quadri}, \& {Rudnick}}]{kriek10}
{Kriek}, M., et al.\ 2010, \apj, 722, L64

\bibitem[{{Kriek} {et~al.}(2009){Kriek}, {van Dokkum}, {Franx}, {Illingworth},
  \& {Magee}}]{kriek09}
{Kriek}, M., {van Dokkum}, P.~G., {Franx}, M., {Illingworth}, G.~D., \&
  {Magee}, D.~K. 2009, \apj, 705, L71

\bibitem[{{Kron}(1980)}]{kron}
{Kron}, R.~G. 1980, \apjs, 43, 305

\bibitem[{{Lawrence} {et~al.}(2007){Lawrence}, {Warren}, {Almaini}, {Edge},
  {Hambly}, {Jameson}, {Lucas}, {Casali}, {Adamson}, {Dye}, {Emerson},
  {Foucaud}, {Hewett}, {Hirst}, {Hodgkin}, {Irwin}, {Lodieu}, {McMahon},
  {Simpson}, {Smail}, {Mortlock}, \& {Folger}}]{lawrence07}
{Lawrence}, A., et al.\ 2007, \mnras, 379, 1599

\bibitem[{{Lee} {et~al.}(2010){Lee}, {Ferguson}, {Somerville}, {Wiklind}, \&
  {Giavalisco}}]{lee10}
{Lee}, S.-K., {Ferguson}, H.~C., {Somerville}, R.~S., {Wiklind}, T., \&
  {Giavalisco}, M. 2010, \apj, 725, 1644

\bibitem[{{Lonsdale} {et~al.}(2003){Lonsdale}, {Smith}, {Rowan-Robinson},
  {Surace}, {Shupe}, {Xu}, {Oliver}, {Padgett}, {Fang}, {Conrow},
  {Franceschini}, {Gautier}, {Griffin}, {Hacking}, {Masci}, {Morrison},
  {O'Linger}, {Owen}, {P{\'e}rez-Fournon}, {Pierre}, {Puetter}, {Stacey},
  {Castro}, {Polletta}, {Farrah}, {Jarrett}, {Frayer}, {Siana}, {Babbedge},
  {Dye}, {Fox}, {Gonzalez-Solares}, {Salaman}, {Berta}, {Condon}, {Dole}, \&
  {Serjeant}}]{lonsdale03}
{Lonsdale}, C.~J., et al.\ 2003, \pasp, 115, 897

\bibitem[{{Lupton} {et~al.}(1999){Lupton}, {Gunn}, \& {Szalay}}]{luptitudes}
{Lupton}, R.~H., {Gunn}, J.~E., \& {Szalay}, A.~S. 1999, \aj, 118, 1406

\bibitem[{{Magorrian} {et~al.}(1998){Magorrian}, {Tremaine}, {Richstone},
  {Bender}, {Bower}, {Dressler}, {Faber}, {Gebhardt}, {Green}, {Grillmair},
  {Kormendy}, \& {Lauer}}]{mag98}
{Magorrian}, J., et al.\ 1998, \aj, 115, 2285

\bibitem[{{Maraston}(2005)}]{maraston05}
{Maraston}, C. 2005, \mnras, 362, 799

\bibitem[{{Maraston} {et~al.}(2010){Maraston}, {Pforr}, {Renzini}, {Daddi},
  {Dickinson}, {Cimatti}, \& {Tonini}}]{maraston10}
{Maraston}, C., {Pforr}, J., {Renzini}, A., {Daddi}, E., {Dickinson}, M.,
  {Cimatti}, A., \& {Tonini}, C. 2010, \mnras, 407, 830

\bibitem[Martig et al.(2009)]{martig09} Martig, M., Bournaud, F., Teyssier, R.\ \& Dekel, A. 2009, \apj, 707, 250

\bibitem[{{McGrath} {et~al.}(2008){McGrath}, {Stockton}, {Canalizo}, {Iye}, \&
  {Maihara}}]{mcgrath08}
{McGrath}, E.~J., {Stockton}, A., {Canalizo}, G., {Iye}, M., \& {Maihara}, T.
  2008, \apj, 682, 303

\bibitem[{{Menanteau} {et~al.}(2001){Menanteau}, {Abraham}, \&
  {Ellis}}]{menanteau01}
{Menanteau}, F., {Abraham}, R.~G., \& {Ellis}, R.~S. 2001, \mnras, 322, 1

\bibitem[{{More} {et~al.}(2009){More}, {van den Bosch}, {Cacciato}, {Mo},
  {Yang}, \& {Li}}]{more09}
{More}, S., {van den Bosch}, F.~C., {Cacciato}, M., {Mo}, H.~J., {Yang}, X., \&
  {Li}, R. 2009, \mnras, 392, 801

\bibitem[{{More} {et~al.}(2011){More}, {van den Bosch}, {Cacciato}, {Skibba},
  {Mo}, \& {Yang}}]{more11}
{More}, S., {van den Bosch}, F.~C., {Cacciato}, M., {Skibba}, R., {Mo}, H.~J.,
  \& {Yang}, X. 2011, \mnras, 410, 210

\bibitem[{{Moster} {et~al.}(2011){Moster}, {Somerville}, {Newman}, \&
  {Rix}}]{moster11}
{Moster}, B.~P., {Somerville}, R.~S., {Newman}, J.~A., \& {Rix}, H.-W. 2011,
  \apj, 731, 113

\bibitem[{{Naab} {et~al.}(2006){Naab}, {Jesseit}, \& {Burkert}}]{naab06}
{Naab}, T., {Jesseit}, R., \& {Burkert}, A. 2006, \mnras, 372, 839

\bibitem[{{Naab} {et~al.}(2007){Naab}, {Johansson}, {Ostriker}, \&
  {Efstathiou}}]{naab07}
{Naab}, T., {Johansson}, P.~H., {Ostriker}, J.~P., \& {Efstathiou}, G. 2007,
  \apj, 658, 710

\bibitem[{{Noeske} {et~al.}(2007){Noeske}, {Weiner}, {Faber}, {Papovich},
  {Koo}, {Somerville}, {Bundy}, {Conselice}, {Newman}, {Schiminovich}, {Le
  Floc'h}, {Coil}, {Rieke}, {Lotz}, {Primack}, {Barmby}, {Cooper}, {Davis},
  {Ellis}, {Fazio}, {Guhathakurta}, {Huang}, {Kassin}, {Martin}, {Phillips},
  {Rich}, {Small}, {Willmer}, \& {Wilson}}]{noeske07_obs}
{Noeske}, K.~G., et al.\ 2007, \apjl, 660, L43

\bibitem[{{Pannella} {et~al.}(2009){Pannella}, {Gabasch}, {Goranova}, {Drory},
  {Hopp}, {Noll}, {Saglia}, {Strazzullo}, \& {Bender}}]{pannella09}
{Pannella}, M., et al.\ 2009, \apj, 701, 787

\bibitem[{{Papovich} {et~al.}(2006){Papovich}, {Moustakas}, {Dickinson}, {Le
  Floc'h}, {Rieke}, {Daddi}, {Alexander}, {Bauer}, {Brandt}, {Dahlen}, {Egami},
  {Eisenhardt}, {Elbaz}, {Ferguson}, {Giavalisco}, {Lucas}, {Mobasher},
  {P{\'e}rez-Gonz{\'a}lez}, {Stutz}, {Rieke}, \& {Yan}}]{papovich06}
{Papovich}, C., et al.\ 2006, \apj, 640, 92

\bibitem[{{Papovich} {et~al.}(2007){Papovich}, {Rudnick}, {Le Floc'h}, {van
  Dokkum}, {Rieke}, {Taylor}, {Armus}, {Gawiser}, {Huang}, {Marcillac}, \&
  {Franx}}]{papovich07}
{Papovich}, C., et al.\ 2007, \apj, 668, 45

\bibitem[{{Papovich} {et~al.}(2012)}]{papovich11}
{Papovich}, C., et al.\ 2012, \apj, 750, 93

\bibitem[Patel et al.(2012)]{patel12} Paten, S.\ G., Holden, B.\ P., Kelson, D.\ D., Franx, M., van der Wel, A., \& Illingworth, G.\ D.\ 2012, \apj, 748, L27

\bibitem[{{Peng} {et~al.}(2002){Peng}, {Ho}, {Impey}, \& {Rix}}]{galfit}
{Peng}, C.~Y., {Ho}, L.~C., {Impey}, C.~D., \& {Rix}, H.-W. 2002, \aj, 124, 266

\bibitem[{{Peng} {et~al.}(2006){Peng}, {Impey}, {Rix}, {Kochanek}, {Keeton},
  {Falco}, {Leh{\'a}r}, \& {McLeod}}]{peng06}
{Peng}, C.~Y., {Impey}, C.~D., {Rix}, H.-W., {Kochanek}, C.~S., {Keeton},
  C.~R., {Falco}, E.~E., {Leh{\'a}r}, J., \& {McLeod}, B.~A. 2006, \apj, 649,
  616

\bibitem[{{Peng} {et~al.}(2011){Peng}, {Lilly}, {Renzini}, \&
  {Carollo}}]{peng11}
{Peng}, Y., {Lilly}, S.~J., {Renzini}, A., \& {Carollo}, M. 2011, arXiv:1106.2546

\bibitem[{{Peng} {et~al.}(2010){Peng}, {Lilly}, {Kova{\v c}}, {Bolzonella},
  {Pozzetti}, {Renzini}, {Zamorani}, {Ilbert}, {Knobel}, {Iovino}, {Maier},
  {Cucciati}, {Tasca}, {Carollo}, {Silverman}, {Kampczyk}, {de Ravel},
  {Sanders}, {Scoville}, {Contini}, {Mainieri}, {Scodeggio}, {Kneib}, {Le
  F{\`e}vre}, {Bardelli}, {Bongiorno}, {Caputi}, {Coppa}, {de la Torre},
  {Franzetti}, {Garilli}, {Lamareille}, {Le Borgne}, {Le Brun}, {Mignoli},
  {Perez Montero}, {Pello}, {Ricciardelli}, {Tanaka}, {Tresse}, {Vergani},
  {Welikala}, {Zucca}, {Oesch}, {Abbas}, {Barnes}, {Bordoloi}, {Bottini},
  {Cappi}, {Cassata}, {Cimatti}, {Fumana}, {Hasinger}, {Koekemoer},
  {Leauthaud}, {Maccagni}, {Marinoni}, {McCracken}, {Memeo}, {Meneux}, {Nair},
  {Porciani}, {Presotto}, \& {Scaramella}}]{peng10}
{Peng}, Y.-J., et al.\ 2010, \apj, 721, 193

\bibitem[{{Prochaska} \& {Hennawi}(2009)}]{prochaska09}
{Prochaska}, J.~X., \& {Hennawi}, J.~F. 2009, \apj, 690, 1558

\bibitem[{{Rix} {et~al.}(2004){Rix}, {Barden}, {Beckwith}, {Bell}, {Borch},
  {Caldwell}, {H{\"a}ussler}, {Jahnke}, {Jogee}, {McIntosh}, {Meisenheimer},
  {Peng}, {Sanchez}, {Somerville}, {Wisotzki}, \& {Wolf}}]{rix04}
{Rix}, H.-W., et al.\ 2004, \apjs, 152, 163

\bibitem[{{Robaina} {et~al.}(2010){Robaina}, {Bell}, {van der Wel},
  {Somerville}, {Skelton}, {McIntosh}, {Meisenheimer}, \& {Wolf}}]{robaina10}
{Robaina}, A.~R., {Bell}, E.~F., {van der Wel}, A., {Somerville}, R.~S.,
  {Skelton}, R.~E., {McIntosh}, D.~H., {Meisenheimer}, K., \& {Wolf}, C. 2010,
  \apj, 719, 844

\bibitem[{{Ruhland} {et~al.}(2009){Ruhland}, {Bell}, {H{\"a}u{\ss}ler},
  {Taylor}, {Barden}, \& {McIntosh}}]{ruhland09}
{Ruhland}, C., {Bell}, E.~F., {H{\"a}u{\ss}ler}, B., {Taylor}, E.~N., {Barden},
  M., \& {McIntosh}, D.~H. 2009, \apj, 695, 1058

\bibitem[{{Salim} {et~al.}(2005){Salim}, {Charlot}, {Rich}, {Kauffmann},
  {Heckman}, {Barlow}, {Bianchi}, {Byun}, {Donas}, {Forster}, {Friedman},
  {Jelinsky}, {Lee}, {Madore}, {Malina}, {Martin}, {Milliard}, {Morrissey},
  {Neff}, {Schiminovich}, {Seibert}, {Siegmund}, {Small}, {Szalay}, {Welsh}, \&
  {Wyder}}]{salim05}
{Salim}, S., et al.\ 2005, \apjl, 619, L39

\bibitem[{{Salim} {et~al.}(2007){Salim}, {Rich}, {Charlot}, {Brinchmann},
  {Johnson}, {Schiminovich}, {Seibert}, {Mallery}, {Heckman}, {Forster},
  {Friedman}, {Martin}, {Morrissey}, {Neff}, {Small}, {Wyder}, {Bianchi},
  {Donas}, {Lee}, {Madore}, {Milliard}, {Szalay}, {Welsh}, \& {Yi}}]{salim07}
{Salim}, S., et al.\ 2007, \apjs, 173, 267

\bibitem[{{Schlegel} {et~al.}(1998){Schlegel}, {Finkbeiner}, \&
  {Davis}}]{sfb98}
{Schlegel}, D.~J., {Finkbeiner}, D.~P., \& {Davis}, M. 1998, \apj, 500, 525

\bibitem[{{Schweizer} \& {Seitzer}(1992)}]{sch92}
{Schweizer}, F., \& {Seitzer}, P. 1992, \aj, 104, 1039

\bibitem[{{Scoville} {et~al.}(2007){Scoville}, {Aussel}, {Brusa}, {Capak},
  {Carollo}, {Elvis}, {Giavalisco}, {Guzzo}, {Hasinger}, {Impey}, {Kneib},
  {LeFevre}, {Lilly}, {Mobasher}, {Renzini}, {Rich}, {Sanders}, {Schinnerer},
  {Schminovich}, {Shopbell}, {Taniguchi}, \& {Tyson}}]{scoville07}
{Scoville}, N., et al.\ 2007, \apjs, 172, 1

\bibitem[{{S\'ersic}(1968)}]{sersic}
{S\'ersic}, J.~L. 1968, {Atlas de galaxias australes} (Cordoba, Argentina:
  Observatorio Astronomico, 1968)

\bibitem[{{Simard} {et~al.}(2011){Simard}, {Mendel}, {Patton}, {Ellison}, \&
  {McConnachie}}]{simard11}
{Simard}, L., {Mendel}, J.~T., {Patton}, D.~R., {Ellison}, S.~L., \&
  {McConnachie}, A.~W. 2011, \apjs, 196, 11

\bibitem[{{Skrutskie} {et~al.}(2006){Skrutskie}, {Cutri}, {Stiening},
  {Weinberg}, {Schneider}, {Carpenter}, {Beichman}, {Capps}, {Chester},
  {Elias}, {Huchra}, {Liebert}, {Lonsdale}, {Monet}, {Price}, {Seitzer},
  {Jarrett}, {Kirkpatrick}, {Gizis}, {Howard}, {Evans}, {Fowler}, {Fullmer},
  {Hurt}, {Light}, {Kopan}, {Marsh}, {McCallon}, {Tam}, {Van Dyk}, \&
  {Wheelock}}]{2mass}
{Skrutskie}, M.~F., et al.\ 2006, \aj, 131, 1163

\bibitem[{{Somerville} {et~al.}(2008){Somerville}, {Hopkins}, {Cox},
  {Robertson}, \& {Hernquist}}]{somerville08}
{Somerville}, R.~S., {Hopkins}, P.~F., {Cox}, T.~J., {Robertson}, B.~E., \&
  {Hernquist}, L. 2008, \mnras, 391, 481

\bibitem[{{Somerville} {et~al.}(2004){Somerville}, {Lee}, {Ferguson},
  {Gardner}, {Moustakas}, \& {Giavalisco}}]{somerville04}
{Somerville}, R.~S., {Lee}, K., {Ferguson}, H.~C., {Gardner}, J.~P.,
  {Moustakas}, L.~A., \& {Giavalisco}, M. 2004, \apj, 600, L171

\bibitem[{{Springel} {et~al.}(2005){Springel}, {Di Matteo}, \&
  {Hernquist}}]{springel05}
{Springel}, V., {Di Matteo}, T., \& {Hernquist}, L. 2005, \apjl, 620, L79

\bibitem[{{Stockton} {et~al.}(2004){Stockton}, {Canalizo}, \&
  {Maihara}}]{stockton04}
{Stockton}, A., {Canalizo}, G., \& {Maihara}, T. 2004, \apj, 605, 37

\bibitem[{{Strateva} {et~al.}(2001){Strateva}, {Ivezi{\'c}}, {Knapp}, \& {et
  al.}}]{strateva01}
{Strateva}, I., {et al.} 2001, \aj,
  122, 1861

\bibitem[{{Szomoru} {et~al.}(2011){Szomoru}, {Franx}, {Bouwens}, {van Dokkum},
  {Labbe}, {Illingworth}, \& {Trenti}}]{szomoru11}
{Szomoru}, D., {Franx}, M., {Bouwens}, R.~J., {van Dokkum}, P.~G., {Labbe}, I.,
  {Illingworth}, G.~D., \& {Trenti}, M. 2011, \apj, 734, L22

\bibitem[{{Tal} {et~al.}(2009){Tal}, {van Dokkum}, {Nelan}, \&
  {Bezanson}}]{tal09}
{Tal}, T., {van Dokkum}, P.~G., {Nelan}, J., \& {Bezanson}, R. 2009, \aj, 138,
  1417

\bibitem[{{Taylor} {et~al.}(2009){Taylor}, {Franx}, {van Dokkum}, {Quadri},
  {Gawiser}, {Bell}, {Barrientos}, {Blanc}, {Castander}, {Damen},
  {Gonzalez-Perez}, {Hall}, {Herrera}, {Hildebrandt}, {Kriek}, {Labb{\'e}},
  {Lira}, {Maza}, {Rudnick}, {Treister}, {Urry}, {Willis}, \&
  {Wuyts}}]{taylor09}
{Taylor}, E.~N., et al.\ 2009, \apjs, 183, 295

\bibitem[{Taylor} {et~al.}(2010)]{taylor10}
{Taylor}, E.~N., Franx, M., Brinchmann, J., van der Wel, A.\ \& van Dokkum, P.\ G., 2010, \apj, 722, 1

\bibitem[{{Toomre} \& {Toomre}(1972)}]{toomre}
{Toomre}, A., \& {Toomre}, J. 1972, \apj, 178, 623

\bibitem[{{Tremonti} {et~al.}(2004){Tremonti}, { }, { }, \& {et
  al.}}]{tremonti04}
{Tremonti}, C.~A., {et al.} 2004, \apj, 613, 898

\bibitem[{{Tremonti} {et~al.}(2007){Tremonti}, {Moustakas}, \&
  {Diamond-Stanic}}]{tremonti07}
{Tremonti}, C.~A., {Moustakas}, J., \& {Diamond-Stanic}, A.~M. 2007, \apjl,
  663, L77

\bibitem[{{van den Bergh}(2009)}]{vandenbergh09}
{van den Bergh}, S. 2009, \apj, 702, 1502

\bibitem[{{van der Wel}(2008)}]{vdw08}
{van der Wel}, A. 2008, \apjl, 675, L13

\bibitem[{{van der Wel} {et~al.}(2010){van der Wel}, {Bell}, {Holden},
  {Skibba}, \& {Rix}}]{vdw10_env}
{van der Wel}, A., {Bell}, E.~F., {Holden}, B.~P., {Skibba}, R.~A., \& {Rix},
  H.-W. 2010, \apj, 714, 1779

\bibitem[{{van der Wel} {et~al.}(2011){van der Wel}, {Rix}, {Wuyts}, {McGrath},
  {Koekemoer}, {Bell}, {Holden}, {Robaina}, \& {McIntosh}}]{vdw11}
{van der Wel}, A., et al.\
  2011, \apj, 730, 38

\bibitem[{{van der Wel} \& {van der Marel}(2008)}]{vdw_vdm08}
{van der Wel}, A., \& {van der Marel}, R.~P. 2008, \apj, 684, 260

\bibitem[{{van Dokkum} {et~al.}(2008){van Dokkum}, {Franx}, {Kriek}, {Holden},
  {Illingworth}, {Magee}, {Bouwens}, {Marchesini}, {Quadri}, {Rudnick},
  {Taylor}, \& {Toft}}]{vd08}
{van Dokkum}, P.~G., et al.\ 2008, \apjl, 677, L5

\bibitem[{van Dokkum} {et~al.}(2011)]{vd11}
{van Dokkum}, P.~G., et al.\ 2011, \apjl, 743, L15

\bibitem[{{Vergani} {et~al.}(2010){Vergani}, {Zamorani}, {Lilly}, {Lamareille},
  {Halliday}, {Scodeggio}, {Vignali}, {Ciliegi}, {Bolzonella}, {Bondi},
  {Kova{\v c}}, {Knobel}, {Zucca}, {Caputi}, {Pozzetti}, {Bardelli}, {Mignoli},
  {Iovino}, {Carollo}, {Contini}, {Kneib}, {Le F{\`e}vre}, {Mainieri},
  {Renzini}, {Bongiorno}, {Coppa}, {Cucciati}, {de la Torre}, {de Ravel},
  {Franzetti}, {Garilli}, {Kampczyk}, {Le Borgne}, {Le Brun}, {Maier}, {Pello},
  {Peng}, {Perez Montero}, {Ricciardelli}, {Silverman}, {Tanaka}, {Tasca},
  {Tresse}, {Abbas}, {Bottini}, {Cappi}, {Cassata}, {Cimatti}, {Guzzo},
  {Koekemoer}, {Leauthaud}, {Maccagni}, {Marinoni}, {McCracken}, {Memeo},
  {Meneux}, {Oesch}, {Porciani}, {Scaramella}, {Capak}, {Sanders}, {Scoville},
  \& {Taniguchi}}]{vergani10}
{Vergani}, D., et al.\ 2010, \aap, 509, A42+

\bibitem[Wake et al.(2012a)]{wake_halo}
  Wake, D.\ A., Franx, M. \& van Dokkum, P.\ G.\ 2012a, submitted to \apj (arXiv:1201.1913)

\bibitem[Wake et al.(2012b)]{wake_quench}
  Wake, D.\ A., van Dokkum, P.\ G.\ \& Franx, M. 2012b, submitted to \apj Letters (arXiv:1201.4998)

\bibitem[{{Warren} {et~al.}(2007){Warren}, {Hambly}, {Dye}, {Almaini}, {Cross},
  {Edge}, {Foucaud}, {Hewett}, {Hodgkin}, {Irwin}, {Jameson}, {Lawrence},
  {Lucas}, {Adamson}, {Bandyopadhyay}, {Bryant}, {Collins}, {Davis}, {Dunlop},
  {Emerson}, {Evans}, {Gonzales-Solares}, {Hirst}, {Jarvis}, {Kendall}, {Kerr},
  {Leggett}, {Lewis}, {Mann}, {McLure}, {McMahon}, {Mortlock}, {Rawlings},
  {Read}, {Riello}, {Simpson}, {Smith}, {Sutorius}, {Targett}, \&
  {Varricatt}}]{warren07}
{Warren}, S.~J., et al.\ 2007, \mnras, 375, 213

\bibitem[{{Weinmann} {et~al.}(2010){Weinmann}, {Kauffmann}, {von der Linden},
  \& {De Lucia}}]{weinmann10}
{Weinmann}, S.~M., {Kauffmann}, G., {von der Linden}, A., \& {De Lucia}, G.
  2010, \mnras, 406, 2249

\bibitem[{{Weinmann} {et~al.}(2011){Weinmann}, {van den Bosch}, \&
  {Pasquali}}]{weinmann11}
{Weinmann}, S.~M., {van den Bosch}, F.~C., \& {Pasquali}, A. 2011, arXiv:1101.3244

\bibitem[{{Whitaker} {et~al.}(2010){Whitaker}, {van Dokkum}, {Brammer},
  {Kriek}, {Franx}, {Labb{\'e}}, {Marchesini}, {Quadri}, {Bezanson},
  {Illingworth}, {Lee}, {Muzzin}, {Rudnick}, \& {Wake}}]{whitaker10}
{Whitaker}, K.~E., et al.\ 2010, \apj, 719, 1715

\bibitem[{Whitaker} {et~al.}(2011)]{whitaker11}
{Whitaker}, K.~E., et al.\ 2011, \apj, 735, 86

\bibitem[{{Williams} {et~al.}(2009){Williams}, {Quadri}, {Franx}, {van Dokkum},
  \& {Labb{\'e}}}]{williams09}
{Williams}, R.~J., {Quadri}, R.~F., {Franx}, M., {van Dokkum}, P., \&
  {Labb{\'e}}, I. 2009, \apj, 691, 1879

\bibitem[{{Williams} {et~al.}(2010)}]{williams10}
{Williams}, R.~J., {Quadri}, R.~F., {Franx}, M., {van Dokkum}, P., Toft, S., Kriek, M. \& {Labb{\'e}}, I. 2010, \apj, 713, 738

\bibitem[{{Wuyts} {et~al.}(2011{\natexlab{a}}){Wuyts}, {F{\"o}rster Schreiber},
  {Lutz}, {Nordon}, {Berta}, {Altieri}, {Andreani}, {Aussel}, {Bongiovanni},
  {Cepa}, {Cimatti}, {Daddi}, {Elbaz}, {Genzel}, {Koekemoer}, {Magnelli},
  {Maiolino}, {McGrath}, {P{\'e}rez Garc{\'{\i}}a}, {Poglitsch}, {Popesso},
  {Pozzi}, {Sanchez-Portal}, {Sturm}, {Tacconi}, \& {Valtchanov}}]{wuyts11sfr}
{Wuyts}, S., et al.\ 2011{\natexlab{a}}, \apj, 738, 106

\bibitem[{{Wuyts} {et~al.}(2011{\natexlab{b}}){Wuyts}, {Forster Schreiber},
  {van der Wel}, {Magnelli}, {Guo}, {Genzel}, {Lutz}, {Aussel}, {Berta},
  {Cava}, {Gracia-Carpio}, {Kocevski}, {Koekemoer}, {Lee}, {Le Floc'h},
  {McGrath}, {Nordon}, {Popesso}, {Pozzi}, {Riguccini}, {Rodighiero},
  {Saintonge}, \& {Tacconi}}]{wuyts11}
{Wuyts}, S., et al.\ 2011{\natexlab{b}}, \apj, 742, 96

\bibitem[{{Wuyts} {et~al.}(2009){Wuyts}, {Franx}, {Cox}, {F{\"o}rster
  Schreiber}, {Hayward}, {Hernquist}, {Hopkins}, {Labb{\'e}}, {Marchesini},
  {Robertson}, {Toft}, \& {van Dokkum}}]{wuyts09}
{Wuyts}, S., et al.\
  2009, \apj, 700, 799

\bibitem[{{Wuyts} {et~al.}(2007){Wuyts}, {Labb{\'e}}, {Franx}, {Rudnick}, {van
  Dokkum}, {Fazio}, {F{\"o}rster Schreiber}, {Huang}, {Moorwood}, {Rix},
  {R{\"o}ttgering}, \& {van der Werf}}]{wuyts07}
{Wuyts}, S., et al.\ 2007,
  \apj, 655, 51

\bibitem[{{Wuyts} {et~al.}(2008){Wuyts}, {Labb{\'e}}, {Schreiber}, {Franx},
  {Rudnick}, {Brammer}, \& {van Dokkum}}]{wuyts08}
{Wuyts}, S., {Labb{\'e}}, I., {Schreiber}, N.~M.~F., {Franx}, M., {Rudnick},
  G., {Brammer}, G.~B., \& {van Dokkum}, P.~G. 2008, \apj, 682, 985

\bibitem[Yang et al.(2009)]{yang09} Yang, X., Mo, H.\ J.\ \& van den Bosch, F.\ C. 2009, \apj, 693, 830

\bibitem[{{Yi} {et~al.}(2005){Yi}, {Yoon}, {Kaviraj}, {Deharveng}, {Rich},
  {Salim}, {Boselli}, {Lee}, {Ree}, {Sohn}, {Rey}, {Lee}, {Rhee}, {Bianchi},
  {Byun}, {Donas}, {Friedman}, {Heckman}, {Jelinsky}, {Madore}, {Malina},
  {Martin}, {Milliard}, {Morrissey}, {Neff}, {Schiminovich}, {Siegmund},
  {Small}, {Szalay}, {Jee}, {Kim}, {Barlow}, {Forster}, {Welsh}, \&
  {Wyder}}]{yi05}
{Yi}, S.~K., et al.\ 2005, \apjl, 619, L111

\bibitem[{{Zheng} {et~al.}(2007){Zheng}, {Bell}, {Papovich}, {Wolf},
  {Meisenheimer}, {Rix}, {Rieke}, \& {Somerville}}]{zheng07}
{Zheng}, X.~Z., {Bell}, E.~F., {Papovich}, C., {Wolf}, C., {Meisenheimer}, K.,
  {Rix}, H.-W., {Rieke}, G.~H., \& {Somerville}, R. 2007, \apjl, 661, L41

\end{thebibliography}
\end{document}